\documentclass[aps, prd, twocolumn, superscriptaddress, preprintnumbers, longbibliography,nofootinbib, 10pt]{revtex4-1}

\usepackage[top=2.8cm, bottom=2.8cm, left=2cm, right=2cm]{geometry}
\usepackage[utf8]{inputenc} 
\usepackage[T1]{fontenc}
\usepackage{amsmath, amssymb, lmodern} 
\usepackage[caption = false]{subfig}
\usepackage{graphicx}
\usepackage{natbib}
\usepackage{booktabs}
\usepackage{color}
\usepackage{hyperref}
\usepackage{tabularray}
\usepackage{comment}
\usepackage{subfig}
\hypersetup{colorlinks, citecolor = blue, filecolor = blue, linkcolor = blue, urlcolor = blue}

\usepackage{soul}
\definecolor{darkgreen}{rgb}{0, 0.6, 0}
\begin{document}

\preprint{PI/UAN-2025-729FT}

\title{Dynamical Shortcomings in the Generalized SU(2) Proca Theory: \\ Challenges for Cosmic Acceleration}

%%%%%%%%%%% AUTHORS INFO %%%%%%%%%%%%%%%%%%%%%%
\author{Santiago Garc\'ia-Serna}
\email{santiago.serna@correounivalle.edu.co}
\affiliation{Departamento  de  F\'isica,  Universidad  del Valle, Ciudad  Universitaria Mel\'endez,  Santiago de Cali  760032,  Colombia}

\author{J. Bayron Orjuela-Quintana}
\email{john.orjuela@correounivalle.edu.co}
\affiliation{Departamento  de  F\'isica,  Universidad  del Valle, Ciudad  Universitaria Mel\'endez,  Santiago de Cali  760032,  Colombia}

\author{Yeinzon Rodr\'{\i}guez}
\email{yeinzon.rodriguez@uan.edu.co}
\affiliation{Centro de Investigaciones en Ciencias B\'asicas y Aplicadas, Universidad Antonio Nari\~no, \\ Cra 3 Este \# 47A-15, Bogot\'a D.C. 110231, Colombia}
\affiliation{Escuela  de  F\'isica,  Universidad  Industrial  de  Santander, \\ Ciudad  Universitaria,  Bucaramanga  680002,  Colombia}

\author{Gabriel G\'omez}
\email{luis.gomezd@umayor.cl}
\affiliation{Centro Multidisciplinario  de  F\'isica, Vicerrector\'{\i}a de Investigaci\'on, Universidad Mayor, \\ Camino La Pir\'amide 5750,  Huechuraba, 8580745,  Santiago, Chile}

\author{C\'esar A. Valenzuela-Toledo}
\email{cesar.valenzuela@correounivalle.edu.co}
\affiliation{Departamento  de  F\'isica,  Universidad  del Valle, Ciudad  Universitaria Mel\'endez,  Santiago de Cali  760032,  Colombia}

%%%%%%%%%%% AUTHORS INFO %%%%%%%%%%%%%%%%%%%%%%

%%%%%%%%%%%%%%%%%%
\begin{abstract}
%%%%%%%%%%%%%%%%%%

The Generalized SU(2) Proca (GSU2P) theory has recently garnered attention for its potential to describe key phases of cosmic evolution, including primordial inflation and late-time accelerated expansion. However, its full cosmological implications remain unexplored. In this work, we perform a comprehensive analysis of the dynamical properties of the GSU2P theory in a flat, homogeneous, and isotropic spacetime, 
%focusing on the conditions under which the theory can drive accelerated expansion. 
through a dynamical-system approach.
%we uncover several critical issues that prevent the vector field from consistently sustaining this acceleration. 
Our analysis reveals the presence of {three pairs of fixed points, one of them} corresponding to de-Sitter expansion which may represent either a stable or unstable phase in the evolution of the universe. 
 {These points, nonetheless, give rise to an indeterminate or infinite Hubble parameter, which renders them cosmologically unviable.} Additionally, we find two key pseudostationary states: the ``attractor lines'', along which the system exhibits constant-roll dynamics, and the ``central zone'', characterized by oscillatory radiation-like behaviour of the field. The dynamics within the central zone could represent a graceful exit from the primordial inflationary phase to a radiation dominated phase, or a state of the dark energy component prior to the late-time cosmic acceleration. However, within the central zone, 
%the vector field density undergoes sign reversals during oscillations, leading 
the dynamics of the vector field leads
to recurrent instances of a nonphysical expansion rate. The absence of a limit cycle in the central zone further exacerbates the issue, as the system may follow unbounded phase-space trajectories, and the expansion rate becomes complex once it escapes the region. 
Collectively, these challenges undermine the viability of the GSU2P theory as a cosmological model for cosmic acceleration. 
%Although the theory can reproduce constant-roll behaviour along the attractor lines, its overall dynamics falls short of providing a consistent framework for either inflation or late-time accelerated expansion.
%Our analysis reveals that the system's fixed points result in non-viable cosmological scenarios, with vector field densities that are either negative or indeterminate, resulting in a nonphysical expansion rate. 
% near the phase space origin.
%Additionally, we identify singularities in the phase space arising from the regularization of the dynamical equations. These singularities further hinder the system's evolution, rendering it non-integrable. 
%the lack of cosmologically viable attractor points, 
%, and the emergence of singularities due to regularization of the dynamical system
% as a model for stable cosmological evolution.
%recurrent sign reversals of the vector field density and the absence of limit cycles within the central zone

%%%%%%%%%%%%%%%%%%
\end{abstract}
%%%%%%%%%%%%%%%%%%

\maketitle

%%%%%%%%%%%%%%%%%%%%%%%%%%%
\section{Introduction} 
\label{Sec: Introduction}
%%%%%%%%%%%%%%%%%%%%%%%%%%%

The standard cosmological model, $\Lambda$CDM, built upon the interplay between a cosmological constant ($\Lambda$) and cold dark matter (CDM), has been remarkably successful in explaining a wide range of cosmological phenomena~\cite{Peebles2020, Planck:2018vyg, DES:2017qwj, Abbott:2018xao, DES:2021wwk, Riess:1998cb, perlmutter:1998np, SupernovaSearchTeam:2004lze, SDSS:2009ocz, Blake:2011en, Aubourg:2014yra, deBernardis:2000sbo, Jaffe:2003it, Planck:2018jri, Planck:2019evm, DES:2018ekb, Planck:2019evm}. However, several observational tensions and anomalies have reminded us of the provisional character of this model~\cite{Perivolaropoulos:2021jda, Abdalla:2022yfr}. Notable among these are the Hubble tension, which reflects a discrepancy in the measured and inferred expansion rates of the universe~\cite{Wong:2019kwg, Riess:2020fzl, Riess:2021jrx}, and the $\sigma_8$ tension, related to the clustering of matter on large scales~\cite{Heymans:2013fya, Joudaki:2016kym, KiDS:2020suj, DES:2021bvc, DES:2021vln, Preston:2023uup, Dalal:2023olq, Li:2023tui}. These challenges have intensified the search for alternatives to the two foundational pillars of $\Lambda$CDM: general relativity (GR) and the cosmological principle~\cite{Bull:2015stt, CANTATA:2021as}.

The cosmological principle, which asserts that the universe is statistically homogeneous and isotropic on large scales, has been a cornerstone of modern cosmology~\cite{Planck:2018vyg}. However, large-scale cosmic flows~\cite{Kashlinsky:2012gy, Watkins:2023rll} and anomalies in the cosmic microwave background (CMB)~\cite{Schwarz:2015cma}, such as low quadrupole and hemispherical asymmetry~\cite{Polastri:2015rda}, challenge this assumption~\cite{Colin:2019opb, Aluri:2022hzs, Hu:2023eyf, Jones:2023ncn}. These observations suggest the possibility of anisotropic or inhomogeneous cosmological evolution~\cite{Battye:2009ze, Perivolaropoulos:2014lua}, motivating the exploration of models that incorporate deviations from perfect isotropy and homogeneity while remaining consistent with current observational constraints~\cite{BeltranAlmeida:2019fou, Guarnizo:2020pkj, Motoa-Manzano:2020mwe, Orjuela-Quintana:2020klr, Gomez:2021jbo, Orjuela-Quintana:2021zoe, Orjuela-Quintana:2022jrg, Garcia-Serna:2023xfw, Gallego:2024gay, Orjuela-Quintana:2024qfn}.

On the other hand, GR, as the prevailing theory of gravity, is grounded in a 4-dimensional spacetime manifold where the Einstein field equations govern gravitational dynamics. Although remarkably robust~\cite{Collett:2018gpf,Will:2018bme}, GR faces theoretical challenges, including its inability to reconcile with quantum mechanics and its reliance on an unexplained cosmological constant to drive the accelerated expansion of the universe~\cite{Weinberg:1988cp, Martin:2012bt}. Lovelock’s theorem, which demonstrates that GR is the only metric theory of gravity in four dimensions with second-order field equations~\cite{Lovelock:1970zsf, Lovelock:1971yv, Lovelock:1972vz}, underscores the rigidity of GR and highlights the need for extensions to incorporate additional degrees of freedom or higher-dimensional frameworks~\cite{Crisostomi:2017ugk}.

One prominent avenue for modifying gravity involves introducing additional dynamical fields. Horndeski’s seminal work on scalar-tensor theories established the most general second-order scalar-tensor framework, now widely known as Horndeski theory~\cite{Horndeski:1974wa} or generalized Galileon theory~\cite{Deffayet:2011gz,Deffayet:2009wt}. This framework has inspired further extensions~\cite{Rodriguez:2017ckc}, including vector-tensor theories such as the generalized Proca (GP) theory~\cite{Tasinato:2014eka,Heisenberg:2014rta, Allys:2015sht, BeltranJimenez:2016rff,Allys:2016jaq,GallegoCadavid:2019zke}, scalar-vector-tensor (SVT) theories~\cite{Heisenberg:2018acv}, and the generalized SU(2) Proca (GSU2P) theory~\cite{GallegoCadavid:2020dho, GallegoCadavid:2022uzn, BeltranJimenez:2016afo, Allys:2016kbq}. These theories allow for richer gravitational dynamics by incorporating vector fields, scalar fields, or combinations thereof, offering new pathways to address the limitations of $\Lambda$CDM and GR.

Among these extensions, the GSU2P theory is particularly intriguing. By introducing a vector field subject to a global SU(2) symmetry in the action, the theory offers a natural framework to explore anisotropic cosmological evolutions and alternative mechanisms for cosmic acceleration. Although the astrophysical and cosmological implications of Horndeski~\cite{Kobayashi:2019hrl,Kreisch:2017uet, Kobayashi:2011nu}, GP~\cite{DeFelice:2020sdq, DeFelice:2016yws, Emami:2016ldl, DeFelice:2016uil, DeFelice:2016cri, Cardona:2023gzq, Heisenberg:2020xak, Gomez:2020sfz, Gomez:2022okq}, and SVT theories~\cite{Heisenberg:2018vsk,Heisenberg:2018mxx, Heisenberg:2018vti, Cardona:2022lcz, Gonzalez-Espinoza:2023qba} have been extensively studied, the GSU2P theory remains relatively underexplored. Existing work has focused on isolated aspects, such as stability issues~\cite{Gomez:2019tbj}, black hole and neutron star solutions~\cite{Martinez:2022wsy,Gomez:2023wei, Martinez:2024gsj}, inflationary scenarios~\cite{Garnica:2021fuu}, and late-time cosmic acceleration~\cite{Rodriguez:2017wkg}, but a comprehensive analysis of its full cosmological implications, including the complete expansion history of the universe, is still lacking.

In this work, we aim to bridge this gap by performing a detailed investigation of the GSU2P theory in a flat, homogeneous, and isotropic background. Specifically, we employ the dynamical systems approach~\cite{Bahamonde:2017ize} to identify the conditions under which the theory can drive cosmic acceleration, either during the early inflationary phase or the late-time accelerated expansion. Our analysis reveals the existence of {a couple of fixed points that represent} de-Sitter expansion, which could be a stable or a transient state of the universe's evolution, and some ``pseudo-stationary'' states at distinct scales in the two-dimensional phase space, representing accelerated expansion and radiation-like behaviour.

In the regime of large field values, the GSU2P model predicts a constant-roll evolution~\cite{Motohashi:2014ppa,Motohashi:2017vdc,Motohashi:2019tyj}, which may correspond to either an inflationary or late-time acceleration phase. Conversely, in the regime of small field values, the system exhibits oscillatory behaviour between two pseudo-stationary states, mimicking a radiation-like fluid. This behaviour could represent either the graceful exit from an inflationary epoch or a prior phase before the onset of late-time acceleration. 

{Our findings, however, indicate that the theory's health is compromised both in the fixed points and in the transition between the phases described in the previous paragraph due to the emergence of a non-physical expansion rate.} Moreover, although the autonomous system can be regularized, this process uncovers singularities that render the system non-integrable, further challenging its viability as a cosmological model.

We will provide a proof of our claims by developing our arguments in the following order. In Section~\ref{Sec: GSU2P}, we present the GSU2P theory within a cosmological context. Section~\ref{Sec: Accelerated Expansion} introduces a dynamical system description of the model, demonstrating that only {two of the fixed points correspond to viable accelerated solutions that, nonetheless, lead to an indeterminate Hubble parameter.} In Section~\ref{Sec: Pseudo States}, we analyze the pseudo-stationary states in the phase space, highlighting the instabilities that hinder the theory’s viability for cosmic acceleration. Finally, in Section~\ref{Sec: Conclusions}, we summarize our findings and discuss their implications.

%%%%%%%%%%%%%%%%%%%%%%%%%%%%%%%%%%%%%%%%%%%%%%%%%%%%%%%%
\section{The Generalized SU(2) Proca Theory}
\label{Sec: GSU2P}
%%%%%%%%%%%%%%%%%%%%%%%%%%%%%%%%%%%%%%%%%%%%%%%%%%%%%%%%

%%%%%%%%%%%%%%%%%%%%%%%%%%%%%%%%%%
\subsection{General Framework}
%%%%%%%%%%%%%%%%%%%%%%%%%%%%%%%%%%

The generalized SU(2) Proca theory considers the dynamics of a vector field belonging to the Lie algebra of the SU(2) group. The corresponding action, as presented in Refs.~\cite{GallegoCadavid:2020dho, GallegoCadavid:2022uzn} (see also Refs.~\cite{BeltranJimenez:2016afo, Allys:2016kbq} for  older constructions and Ref.~\cite{GallegoCadavid:2021ljh} for an extended version), is designed to respect global invariance under this group of transformations and to propagate the right number of degrees of freedom~\cite{ErrastiDiez:2019trb} (see, anyway, Refs.~\cite{ErrastiDiez:2023gme, Janaun:2023nxz}), thereby circumventing Ostrogradski instabilities~\cite{Ostrogradsky:1850fid, Woodard:2006nt, Woodard:2015zca}. The action of the GSU2P theory is given by:
\begin{align}
\label{Eq: Action GSU2P}
S \equiv \int \text{d}x^4 \sqrt{-g} & \left\{ \mathcal{L}_{\text{EH}} + \mathcal{L}_{\text{YM}} + \sum_{i=1}^{2} \chi_i \mathcal{L}_2^i \right. \nonumber \\
 &\left.+ \sum_{i=3}^{7} \frac{\chi_i}{m_{\text{P}}^2} \mathcal{L}_{2}^i + \sum_{i=1}^{6} \frac{\alpha_i}{m_{\text{P}}^2} \mathcal{L}_{4,2}^i \right\},    
\end{align}
where
\begin{equation}
    \mathcal{L}_{\text{EH}} \equiv \frac{m_{\text{P}}^2}{2}R, \qquad
    \mathcal{L}_{\text{YM}} \equiv-\frac{1}{4}F^a_{\ \mu\nu}F_a^{\ \mu\nu},
\end{equation}
denote the Einstein-Hilbert and the Yang-Mills Lagrangians, respectively, while the different vector-tensor interactions are:
\begin{align}
    \mathcal{L}_2^1 &\equiv B^a_{\ \mu} B_a^{\ \mu} B^b_{\ \nu} B_b^{\ \nu}, \\
    %%%%%%%%%%%%
    \mathcal{L}_2^2 &\equiv B^a_{\ \mu} B^{b\mu} B_a^{\ \nu} B_{b\nu}, \\
    %%%%%%%%%%%%
    \mathcal{L}_2^3 &\equiv A_a^{\ \mu \nu} A^\rho{ }_\nu{ }^a B_{\ \mu}^b B_{\rho b}, \\
    %%%%%%%%%%%%
    \mathcal{L}_2^4 &\equiv A_a^{\ \mu \nu} A^\rho{ }_\nu{ }^b B_{\mu b} B_{\ \rho}^a, \\
    %%%%%%%%%%%%
    \mathcal{L}_2^5 &\equiv A_a^{\ \mu \nu} A^\rho{ }_\nu{ }^b B_{\ \mu}^a B_{b\rho}, \\
    %%%%%%%%%%%%
    \mathcal{L}_2^6 &\equiv A_a^{\ \mu \nu} A_{\ \mu \nu}^a{ } B_b^{\ \rho} B^b_{\ \rho}, \\
    %%%%%%%%%%%%
    \mathcal{L}_2^7 &\equiv A_a^{\mu \nu} A_{\mu \nu}^b B^a_{\ \rho} B_b^{\ \rho},
\end{align}
and
\begin{align}
    \mathcal{L}_{4, 2}^1 & \equiv B_b^{\ \rho} B^b_{\ \rho} \left[S_{\ \ \mu}^{a\mu} S_{a \nu}^{\ \ \nu} - S_{\ \ \nu}^{a \mu} S_{\mu a}^{\ \ \nu} \right] \nonumber \\
    &+ 2 B_a^{\ \rho} B_{b\rho} \left[ S_{\ \ \mu}^{a\mu} S_{\ \ \nu}^{b\nu} - S_{\ \ \nu}^{a\mu} S_{\ \ \mu}^{b\nu}\right], \\
    %%%%%%%%%%%%
    \mathcal{L}_{4,2}^2 & \equiv A_{\ \mu \nu}^a S_{\ \ \sigma}^{b\mu} B_a^{\ \nu} B_b^{\ \sigma} - A_{\ \mu \nu}^a S_{\ \ \sigma}^{b\mu} B_b^{\ \nu} B_a^{\ \sigma} \nonumber \\
    &+ A_{\ \mu \nu}^a S_{\ \ \rho}^{b\rho} B_a^{\ \mu} B_b^{\ \nu}, \\
    %%%%%%%%%%%%
    \mathcal{L}_{4,2}^3 & \equiv B^{\mu a} R^\alpha{ }_{\sigma \rho \mu} B_{\alpha a} B^{\rho b} B_b^{\ \sigma} \nonumber \\
    &+ \frac{3}{4} B_b^{\ \mu} B^b_{\ \mu} B^a_{\ \nu} B_a^{\ \nu} R, \\
    %%%%%%%%%%%%
    \mathcal{L}_{4,2}^4 & \equiv \left(B_b^{\ \mu} B^b_{\ \mu} B^a_{\ \nu} B_a^{\ \nu} + 2 B_a^{\ \mu} B_{b\mu} B^a_{\ \nu} B^{b\nu}\right) R, \\
    %%%%%%%%%%%%
    \mathcal{L}_{4,2}^5 & \equiv G_{\mu \nu} B^{a\mu} B_a^{\ \nu} B^b_{\ \rho} B_b^{\ \rho}, \\
    %%%%%%%%%%%%
    \mathcal{L}_{4,2}^6  &\equiv G_{\mu \nu} B^{a\mu} B^{b\nu} B_a^{\ \rho} B_{b\rho}.
\end{align}
In the previous expressions, $g$ is the determinant of the metric, $m_{\text{P}}$ is the reduced Planck mass, $R$ is the Ricci scalar, $G_{\mu\nu}$ is the Einstein tensor, $R^{\alpha}_{\ \sigma\rho\mu}$ is the Riemann tensor, $\epsilon_{abc}$ is the Levi-Civita symbol, $B^a_{\ \mu}$ is the vector field within the Lie algebraic structure of SU(2) whose strength tensor is $F^{a}_{\ \mu\nu}\equiv\nabla_{\mu}B^a_{\ \nu}-\nabla_{\nu}B^a_{\ \mu}+\tilde{g}{\epsilon^a}_{bc}B^{b}_{\ \mu}B^{c}_{\ \nu}$, where $\tilde{g}$ is the SU(2) coupling constant, and we define the symmetric and antisymmetric tensors $S^{a}_{\ \mu\nu}\equiv\nabla_{\mu}B^a_{\ \nu}+\nabla_{\nu}B^a_{\ \mu}$ and $A^{a}_{\ \mu\nu}\equiv\nabla_{\mu}B^a_{\ \nu}-\nabla_{\nu}B^a_{\ \mu}$, respectively. All the $\alpha_i$ and $\chi_i$ are arbitrary dimensionless constants. Hereinafter, Greek indices denote space-time indices that run from $0$ to $3$ while Latin indices run from $1$ to $3$ and denote space indices and/or SU(2) group indices. 

It is worth mentioning that the only pieces of the GSU2P theory that have been considered are those that are relevant for the cosmic acceleration mechanism discussed in Refs.~\cite{Garnica:2021fuu, Rodriguez:2017wkg}, i.e., those that, prior to covariantization, involve two derivatives and two vector fields, or four vector fields (see Refs.~\cite{GallegoCadavid:2020dho, GallegoCadavid:2022uzn}).

%%%%%%%%%%%%%%%%%%%%%%%%%%%%%%%%%%%%%%%%%%%%%%%%%%%%%%%%%%%
\subsection{Stability Conditions and Gravitational Wave-Speed Constraint}
%%%%%%%%%%%%%%%%%%%%%%%%%%%%%%%%%%%%%%%%%%%%%%%%%%%%%%%%%%%

The action defined in Eq.~\eqref{Eq: Action GSU2P} has been meticulously constructed to circumvent Ostrogradski's instability, thereby ensuring the correct number of propagating degrees of freedom~\cite{ErrastiDiez:2019trb} (see, however, Refs.~\cite{ErrastiDiez:2023gme, Janaun:2023nxz}). Nonetheless, for the theory to be considered physically viable, it is imperative that it remains free from other forms of instabilities which could undermine its consistency. Notable pathologies that must be avoided include ghost instabilities and gradient or Laplacian instabilities.

Ghost instabilities manifest when the linearized perturbations exhibit negative kinetic energy terms, resulting in nonphysical behaviour when the ghost field interacts with other fields. In contrast, Laplacian instabilities occur when the propagation speed of perturbations is imaginary, leading to the uncontrollable, often exponential, growth of initially small perturbations~\cite{Sbisa:2014pzo}. Furthermore, the detection of gravitational waves (GW) by LIGO~\cite{LIGOScientific:2017vwq} and the subsequent determination of their speed, which has been shown to be equal to the speed of light with astonishing precision~\cite{Liu:2020slm, Baker:2022eiz}, have imposed stringent constraints on several modified gravity theories~\cite{Ezquiaga:2017ekz, Sakstein:2017xjx, Creminelli:2017sry, Kreisch:2017uet, Baker:2017hug, Jana:2018djs}.

In the context of the GSU2P theory, it has been demonstrated that to prevent ghost and Laplacian instabilities from appearing and to ensure tensor modes to propagate at the speed of light, the parameters in the action \eqref{Eq: Action GSU2P} must satisfy the following conditions~\cite{Garnica:2021fuu}:
\begin{align}
    \chi_3 &= 0, \\
    \chi_7 &= 5 \alpha_1+\alpha_3-\frac{1}{2}\chi_4 -3 \chi_6, \\
    \alpha_2 &= 2 \alpha_3, \\
    \alpha_4 &= -2 \alpha_1+\frac{7}{20} \alpha_3, \\
    \alpha_5 &= \frac{14}{3} \alpha_3 - \frac{20}{3} \alpha_1, \\
    \alpha_6 &= -20 \alpha_1+6 \alpha_3-3 \alpha_5.
\end{align}
Thus, the final form of the GSU2P theory that satisfies these stability requirements is given by:
\begin{align}
\label{Eq: Stable Action}
S &= \int \text{d}^4 x \, \sqrt{-g} \Bigg[ \mathcal{L}_\text{EH} + \mathcal{L}_\text{YM} + \chi_1 \mathcal{L}^1_2 + \chi_2 \mathcal{L}^2_2 \\
 &\left.+ \frac{\chi_4}{m_\text{P}^2} \left( \mathcal{L}_2^4 - \frac{\mathcal{L}_2^7}{2} \right) + \frac{\chi_5}{m_\text{P}^2}  \mathcal{L}_2^5 + \frac{\chi_6}{m_\text{P}^2} \left( \mathcal{L}_2^6 - 3 \mathcal{L}_2^7 \right) \nonumber \right. \\
 &\left.+ \frac{\alpha_1}{m_\text{P}^2} \left(\mathcal{L}_{4,2}^1- 2 \mathcal{L}_{4,2}^4 - \frac{20}{3} \mathcal{L}_{4,2}^5 + 5 \mathcal{L}_2^7 \right) \nonumber \right.\\
 & + \frac{\alpha_3}{m_\text{P}^2} \left(2 \mathcal{L}_{4,2}^2 + \mathcal{L}_{4,2}^3 + \frac{7}{20} \mathcal{L}_{4,2}^4 + \frac{14}{3} \mathcal{L}_{4,2}^5 - 8 \mathcal{L}_{4,2}^6 + \mathcal{L}_2^7 \right)  \Bigg]. \nonumber
\end{align}
In the following sections, we will focus on the cosmological dynamics encoded in this final action, exploring its implications for the evolution of the universe.

%%%%%%%%%%%%%%%%%%%%%%%%%%%%%%%%%%%%%%%%%%%%%%%%%%%%%%%
\subsection{Homogeneous and Isotropic Configuration}
%%%%%%%%%%%%%%%%%%%%%%%%%%%%%%%%%%%%%%%%%%%%%%%%%%%%%%%

Observational evidence has pointed out that the universe is largely homogeneous and isotropic on cosmological scales~\cite{Planck:2018vyg}.\footnote{There exists controversy around this point~\cite{Aluri:2022hzs}. However, we will neglect a possible anisotropic expansion as a first approximation.} This allows one to describe the geometry of the universe by the flat Friedman-Lema\^itre-Robertson-Walker (FLRW) metric, which in Cartesian coordinates reads:
\begin{equation}
\label{Eq: FLRW metric}
    \text{d}s^2=-\text{d}t^2 + a^2(t) \delta_{i j} \text{d}x^i \text{d} x^j,
\end{equation}
where $a(t)$ is the scale factor, $t$ is the cosmic time, and $x^i$ denotes spatial coordinates. 

The symmetries of this metric significantly constrain the dynamics of cosmological fields. For instance, vector fields inherently break rotational invariance, which can potentially introduce substantial anisotropy into the universe's expansion dynamics. In theories involving vector fields that are not subject to any internal global symmetry in the action, this issue can be addressed either by introducing three identical and orthogonal vector fields, known as the \emph{cosmic triad}~\cite{Armendariz-Picon:2004say, Emami:2016ldl}, or by restricting consideration to time-like vector fields (see, e.g., Refs.~\cite{Koivisto:2008xf, DeFelice:2016yws}). 

In contrast, within the framework of the GSU2P theory, rotational invariance can be preserved by compensating for spatial rotations of the vector field with internal rotations in the isospin space. The most general configuration of the vector field consistent with spatial isotropy has been demonstrated to be given by~\cite{Witten:1976ck, Forgacs:1979zs, Sivers:1986kq}:
\begin{align}
    B_{0a} &= b_0(t)\hat{\bar{r}}_a, \\
    B_{ia} &= b_1(t)\hat{r}_i\hat{\bar{r}}_a+b_2(t)[\delta_{ia}-\hat{r}_i\hat{\bar{r}}_a]+b_3(t)\epsilon_{ia}{}^k\hat{r}_k,
\end{align}
where $b_0(t)$, $b_1(t)$, $b_2(t)$, and $b_3(t)$ are arbitrary functions of time only, $\hat{\bar{r}}$ is the unit vector in the isospin space pointing in the direction of $\vec{B}_i$, and $\hat{r}$ is the unit vector in physical space pointing in the direction of $\vec{B}_a$. The cosmic triad can be naturally accommodated within this general configuration by assuming $b_0(t) = b_3(t) = 0$ and $b_1(t) = b_2(t)$, such that:
\begin{equation}
\label{Eq: triad cosmic}
    B_{0a}(t)=0, \qquad B_{ia}(t) = a(t)\psi(t)\delta_{ia},
\end{equation}
where $b_2 (t) \equiv a(t) \psi(t)$, $\psi(t)$ being the norm of the physical 3D vector fields. 

%%%%%%%%%%%%%%%%%%%%%%%%%%%%%%%%%%%%%%%%%%%%%%%%%%%%%%%%%%
\subsection{Cosmological Dynamics in the GSU2P Theory}
%%%%%%%%%%%%%%%%%%%%%%%%%%%%%%%%%%%%%%%%%%%%%%%%%%%%%%%%%%

The cosmological dynamics encoded in the action~\eqref{Eq: Stable Action} can be revealed through the application of the variational principle. By varying this action with respect to the metric, we derive the gravitational field equations:
\begin{equation}
\label{Eq: Eisntein equation}
    m_\text{P}^2 G_{\mu \nu} = T_{\mu\nu}^{(B)} + T_{\mu\nu}^{(m)},
\end{equation}
where $T_{\mu\nu}^{(B)}$ represents the energy-momentum tensor containing contributions from the SU(2) vector field $B^a_{\ \mu}$, and $T_{\mu\nu}^{(m)}$ is the energy-momentum tensor for the rest of matter fluids in the cosmic budget.

Due to its length, we refrain from presenting the full form of $T_{\mu\nu}^{(B)}$ here.\footnote{The full calculation is available in a \texttt{Mathematica} notebook accessible on the \texttt{GitHub} repository: \href{https://github.com/sagaser/GSU2P}{sagaser/GSU2P}. See also Ref.~\cite{Martinez:2022wsy}.} Substituting the FLRW metric from Eq.~\eqref{Eq: FLRW metric} and the cosmic triad configuration from Eq.~\eqref{Eq: triad cosmic} into the gravitational field equations yields the Friedman equations:
\begin{align}
    3m_{\text{P}}^2H^2 &= \rho_{B} + \rho_m, \label{Eq: First Friedman} \\
    -2m_{\text{P}}^2\dot{H} &= p_{B} + \rho_{B} + \rho_m, \label{Eq: Second Friedman} 
\end{align}
where we have assumed a pressure-less matter fluid with density $\rho_m$. The density $\rho_B$ and pressure $p_B$, which consider contributions from the vector field, are given by:
\begin{align}
    \rho_B &\equiv \left(\dot{\psi} + H \psi \right)^2 \left[ \frac{3}{2} - 9 c_2 \frac{\psi^2}{m_\text{P}^2} \right] + \frac{3}{2} \hat{g}^2\psi^4 \label{Eq: Density} \\
    &+ 6H (c_1 - c_2) \frac{\psi^3 \dot{\psi}}{m_\text{P}^2}, \nonumber \\
    p_B &\equiv \left(\dot{\psi} + H \psi \right)^2\left[\frac{1}{2} + 3 c_2 \frac{\psi^2}{m_\text{P}^2}\right] + \frac{1}{2} \hat{g}^2\psi^4 \label{Eq: Pressure} \\
    &+ 6 \frac{\psi^2}{m_\text{P}^2} (c_2 - c_1)\left\{\dot{\psi}^2-\psi^2\left(H^2+\frac{\dot{H}}{3}\right)+\frac{1}{3}\psi\ddot{\psi}\right\}. \nonumber
\end{align}
In these expressions, the Hubble parameter defined as $H \equiv \dot{a}/a$ represents the expansion rate of the universe, and an over-dot denotes differentiation with respect to cosmic time. The new constants $c_1$ and $c_2$ are defined through:
\begin{equation}
\label{Eq: New Constants}
    \alpha_3 \equiv \alpha_1 + \frac{1}{20}(c_2 - c_1), \quad \chi_5 \equiv - 2 \alpha_1 + \frac{1}{10}(c_1 + 9c_2),
\end{equation}
and we have defined a generalized version of the SU(2) coupling constant as:\footnote{In Ref. \cite{Garnica:2021fuu}, $\hat{g}^2$ was assumed to be positive.  Here, we have relaxed such an assumption.}
\begin{equation}
\label{Eq: Generalized Charge}
    \hat{g}^2 \equiv \tilde{g}^2 - 6\chi_1 - 2\chi_2.
\end{equation}
Finally, varying the action in Eq.~\eqref{Eq: Stable Action} with respect to $B^{a}_{\ \mu}$, and substituting the homogeneous and isotropic configurations for the metric and the field, yields the equation of motion for the sole dynamical degree of freedom representing the vector field:
\begin{align}
    0 &= \ddot{\psi} + 3H \dot{\psi} + \psi\left(2 H^2 + \dot{H} - 6 c_2 \frac{\dot{\psi}^2}{m_\text{P}^2}\right) \nonumber  \\ 
      &+ 2 \psi^3\left[\hat{g}^2+3\left(c_1-2 c_2\right) \frac{H^2}{m_\text{P}^2}+\left(c_1-4 c_2\right) \frac{\dot{H}}{m_\text{P}^2} \right] \nonumber \\
      &-6 c_2 \frac{\psi^2}{m_\text{P}^2} \left(\ddot{\psi} + 3 H \dot{\psi}\right). \label{Eq: Vector EoM}
\end{align}
In the subsequent section, we examine the dynamics of the universe's accelerated expansion as influenced by the vector field.

%%%%%%%%%%%%%%%%%%%%%%%%%%%%%%%%%%%%%%%%%%%%%%%%%%%%%%%%%%%%%%%%%%
\section{Accelerated Expansion Driven by the SU(2) Vector Field}
\label{Sec: Accelerated Expansion}
%%%%%%%%%%%%%%%%%%%%%%%%%%%%%%%%%%%%%%%%%%%%%%%%%%%%%%%%%%%%%%%%%%

According to the current cosmological paradigm, the universe has experienced two phases of accelerated expansion: an early inflationary phase preceding the radiation-dominated epoch and the present accelerated expansion, likely driven by dark energy. To gain insight into the system's asymptotic behaviour, we initially neglect the matter sector, thereby isolating the dynamics of the vector field. This allows us to determine the conditions under which the vector field can induce accelerated expansion, whether in the primordial or late-time phases.

%%%%%%%%%%%%%%%%%%%%%%%%%%%%%%%%%%%%%%%%%%
\subsection{Autonomous System}
%%%%%%%%%%%%%%%%%%%%%%%%%%%%%%%%%%%%%%%%%%

To identify the conditions under which the vector field drives accelerated expansion, it is necessary to determine the parameter space where such solutions exist and assess their stability properties. The asymptotic behaviour of the model, encoded in its fixed points~\cite{Bahamonde:2017ize}, provides valuable insights. To facilitate this analysis, we reformulate the dynamical equations using the following dimensionless variables:
\begin{equation}
    x \equiv \frac{\dot{\psi}}{\sqrt{2} m_{\text{P}}H}, \quad y \equiv \frac{\psi}{\sqrt{2} m_{\text{P}}}, \quad z \equiv \sqrt{\frac{\hat{g}}{2m_\text{P} H}}\psi. \label{Eq: Dynamical System}
\end{equation}
Neglecting $\rho_m$, the first Friedman equation in Eq.~\eqref{Eq: First Friedman} simplifies to the following constraint:\footnote{Since the analysis spans the inflationary epoch in the distant past and the dark energy-dominated future, the effects of radiation are negligible during both periods.}
\begin{equation}
\label{Eq: Friedman Constraint}
    1 = (x + y)^2(1 - 12 c_2 y^2) + 8 (c_1 - c_2) x y^3 + 2 z^4, 
\end{equation}
which allows us to eliminate the variable $z$ from the dynamical system, expressing it in terms of the variables $x$ and $y$. Here, $x$ and $y$  unambiguously represent the speed and the magnitude of the vector field, respectively.

In terms of these variables, the evolution equations of the model reduce to an autonomous system governed by the following set of first-order differential equations:
\begin{align}
\label{Eq: Autonomous set}
x' &= \frac{p}{\sqrt{2}} + x\epsilon, \quad
y' = x,
\end{align}
where the prime denotes differentiation with respect to the number of $e$-folds, $N$, defined as $\text{d}N \equiv H \text{d}t$. The variables $p$ and $\epsilon$ are defined as:
\begin{align}
    p\equiv\frac{\Ddot{\psi}}{m_{\text{P}}H},\quad\epsilon\equiv-\frac{\dot{H}}{H^2},
\end{align}
with $p$ obtained from the vector field equation of motion in Eq.~\eqref{Eq: Vector EoM}, and $\epsilon$ from the second Friedman equation in Eq.~\eqref{Eq: Second Friedman}, which in terms of the new variables read: 
\begin{align}
    \epsilon &= 2 + 12c_1 y^4 - 4y^3(c_1 - c_2)\left( \frac{p}{\sqrt{2}} + \epsilon y \right) \nonumber \\ 
    &- 4(c_1 - 7c_2)xy^3 - 12 (c_1 - 2c_2)x^2 y^2, \label{Eq: epsilon}\\
    \frac{p}{\sqrt{2}} &= 2 y^2 \left(2 x (4 c_1-7 c_2)+3 \sqrt{2} c_2 p\right) + \frac{2 \left(x^2-1\right)}{y} + x \notag \\
    &+ y\left(\epsilon -12 c_2 x^2\right) + 4 y^3 (c_1 \epsilon -3 c_1-4 c_2 \epsilon) \label{Eq: p}.
\end{align}
In these expressions, notice that  $\epsilon$ and $p$ are linearly coupled, allowing them to be expressed solely in terms of the variables $x$ and $y$.

%%%%%%%%%%%%%%%%%%%%%%%%%%%%%%%%%%%%%%%%%%%%%%%%%%%%%%%%%%
\subsection{Fixed Points as Accelerated Solutions}
%%%%%%%%%%%%%%%%%%%%%%%%%%%%%%%%%%%%%%%%%%%%%%%%%%%%%%%%%%

\begin{figure*}[t!]
\centering    
{\includegraphics[width=0.47\textwidth]{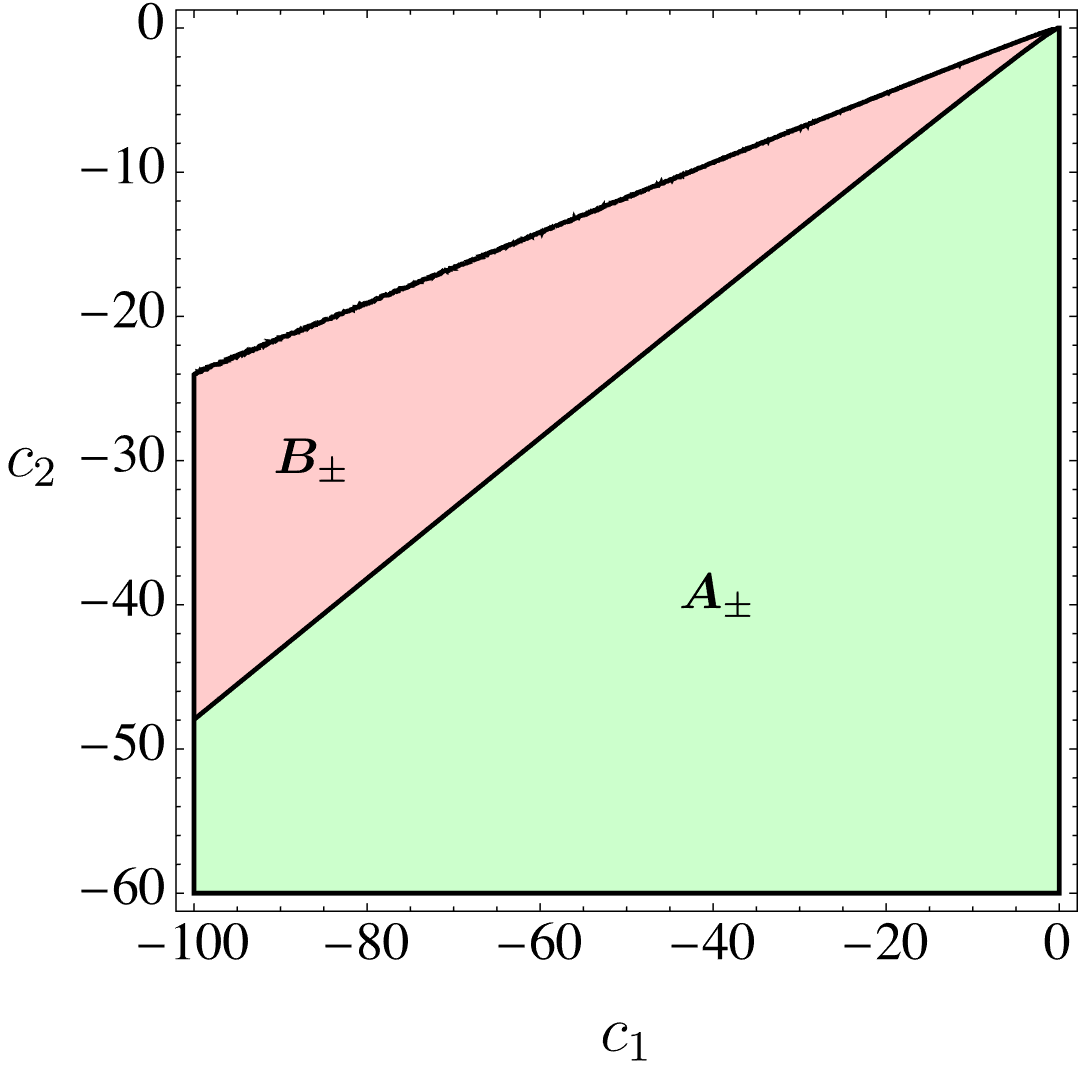}} \hfill
{\includegraphics[width=0.48\textwidth]{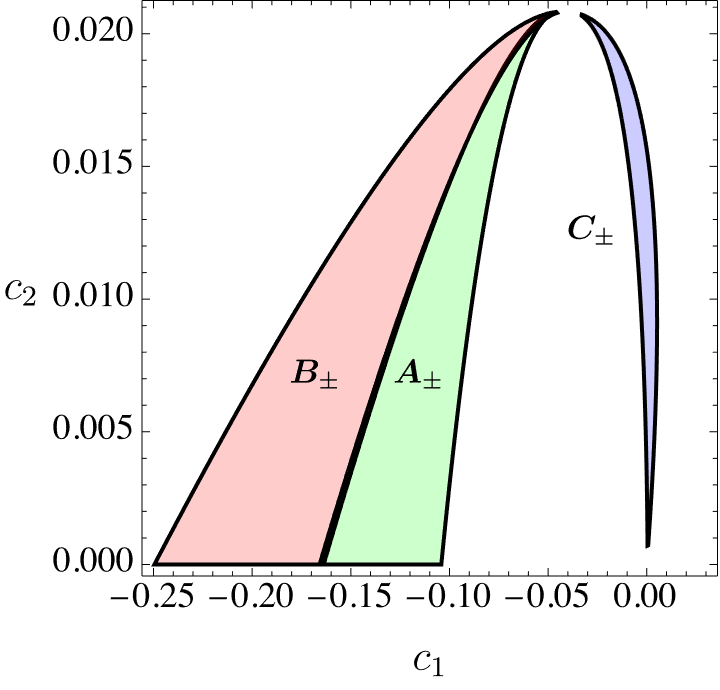}}
\caption{(Left) The regions in the parameter space $\{c_1, c_2\}$ where the fixed points $A_\pm$ (light green) and $B_\pm$ (light red) act as attractors are shown, with the two sets separated by a corresponding bifurcation curve and the attractors representing accelerated expansion. (Right) For small values of the parameters $c_1$ and $c_2$, all fixed points can serve as attractors representing accelerated expansion. However, it is important to note that these regions are considerably smaller compared to those in the left panel.}
\label{Fig: Stability}
\end{figure*}

The fixed points of the system correspond to its stationary states, i.e., where $x' =  y' = 0$ in Eq.~\eqref{Eq: Autonomous set}. Solving the resulting algebraic equations yields the following fixed points:
\begin{align}
A_\pm &= \left\{x\to 0,\,  y\to \pm \sqrt[4]{-\frac{1}{6c_1}} \right\}, \\
B_\pm &= \left\{x\to 0,\,  y\to \pm \frac{1}{2}\sqrt{\frac{1}{6c_2} - \frac{\sqrt{1 - 48c_2}}{6c_2}}\right\}, \\ 
C_\pm &= \left\{x\to 0,\,  y\to \pm\frac{1}{2}\sqrt{\frac{1}{6c_2} + \frac{\sqrt{1 - 48c_2}}{6c_2}}\right\}. 
\end{align}
From these expressions, which only depend on the constants $c_1$ and $c_2$, we observe that the fixed points $A_\pm$, $B_\pm$, and $C_\pm$ assume real values under the following conditions:
\begin{align}
    A_\pm \in \mathbb{R}&: c_1 < 0, \label{Eq: Existence A} \\
    B_\pm \in \mathbb{R}&: c_2 < 0 \quad \lor \quad 0 < c_2 \leq 1/48, \\
    C_\pm \in \mathbb{R}&: 0 < c_2 \leq 1/48. \label{Eq: Existence C} 
\end{align}
For the vector field to drive an accelerated expansion phase, we must ensure that its equation of state $w_B$ satisfies:
\begin{equation}
    w_B \equiv \frac{p_B}{\rho_B} < -\frac{1}{3}.
\end{equation}
At the fixed points $A_\pm$, we find $w_B = -1$, indicating that these solutions correspond to de-Sitter points, where the universe undergoes exponential expansion. Evaluating $w_B$ at the fixed points $B_\pm$ shows that these correspond to accelerated expansion solutions as long as:
\begin{align} 
    B_\pm: & \ 1 + 8 c_1 + \sqrt{1 - 48 c_2} > 32 c_2 \notag \\ 
    \land & \ 1 + 16 c_1 + \sqrt{1 - 48 c_2} < 16 c_2,\label{Eq: Accelerated B} 
\end{align}
while for $C_\pm$, the condition for acceleration is: 
\begin{align}
    C_\pm: & \ \sqrt{1 - 48 c_2} + 16c_2 < 1 + 16 c_1 \notag \\ 
    \land & \ \sqrt{1 - 48 c_2} + 32 c_2  > 1 + 8 c_1. \label{Eq: Accelerated C}
\end{align} \\ 

The solutions reveal distinct cosmological phases associated with different values of the parameters \( c_1 \) and \( c_2 \). Notably, the fixed points \( A_\pm \) represent de-Sitter solutions corresponding to an exponentially expanding universe, while \( B_\pm \) and \( C_\pm \) describe more complex scenarios where the conditions for accelerated expansion are fulfilled under specific parameter ranges, which will be further studied in the next sections. 

%%%%%%%%%%%%%%%%%%%%%%%%%%%%%%%%%%%%%
\subsection{Stability Analysis}
%%%%%%%%%%%%%%%%%%%%%%%%%%%%%%%%%%%%%

Having established the conditions for the fixed points to provide accelerated expansion solutions, we now turn to their stability properties to establish the system's asymptotic behaviour.

As a first approximation, the stability of the fixed points can be determined by analyzing the eigenvalues of the Jacobian matrix, defined as:\footnote{Other techniques are required when the linear stability fails, such as Lyapunov exponents~\cite{Bahamonde:2017ize} or the Malkin's criterion~\cite{Malkin_52}.}
\begin{equation}
J_{ij} \equiv \frac{\text{d} f_i}{\text{d}x_j},
\end{equation}
where $f_i \equiv x'_i$ represents the derivatives of the dynamical variables (i.e., $x'$ and $y'$), and $x_j$ represents each dynamical variable (i.e., $x$ and $y$). The stability of a given fixed point is characterized by the sign of the real part of the eigenvalues of $J_{ij}$, evaluated at the fixed point. Since the Jacobian matrix is, in this case, a $2\times2$ matrix, only two eigenvalues, denoted as $\lambda_1$ and $\lambda_2$, are expected. A fixed point is classified as:
\begin{itemize}
    \item A repeller (or source) if $\text{Re}\{\lambda_1, \lambda_2\} > 0$.
    \item A saddle if the real part of one eigenvalue is positive whereas the real part of the other one is negative.
    \item An attractor (or sink) if $\text{Re}\{\lambda_1, \lambda_2\} < 0$.
\end{itemize}
The stability of the fixed points is crucial for determining the cosmological dynamics. As mentioned before, the universe has undergone two distinct accelerated phases: primordial inflation and late-time dark energy domination. To resolve the flatness, horizon, and unwanted relics problems of classical cosmology, inflation must last for a minimum of 60 $e$-folds, followed by a reheating phase that transitions the universe into the standard Big Bang evolution. Given that inflation represents a transient period of accelerated expansion, it is expected, from a dynamical systems perspective, that inflationary solutions correspond to either a source or a saddle point. In contrast, the current accelerated expansion might persist indefinitely, implying that the corresponding solution might be an attractor point. %ensuring the stability of the present cosmological state.

In our analysis, the eigenvalues of the Jacobian matrix can be computed analytically; however, the resulting expressions are too lengthy and complex to allow for an analytical description of the stability properties of the fixed points in the parameter space. As a result, we adopt the following strategy: we select a representative sector of the parameter space $\{c_1, c_2\}$ and numerically evaluate the eigenvalues, respecting the existence conditions in Eqs.~\eqref{Eq: Existence A}–\eqref{Eq: Existence C}, as well as the conditions for accelerated expansion given in Eqs.~\eqref{Eq: Accelerated B} and \eqref{Eq: Accelerated C}. As demonstrated in subsequent sections, this representative region is sufficiently large to capture the essential dynamics of the system at the fixed points.

The results are presented in Fig.~\ref{Fig: Stability}. In the left panel, we display the regions where $A_\pm$ and $B_\pm$ are accelerated attractors, visibly separated by a bifurcation curve. In the right panel, we show smaller regions where $A_\pm$ and $B_\pm$ are attractors as well, along with the narrow region where $C_\pm$ serves as an attractor.

%%%%%%%%%%%%%%%%%%%%%%%%%%%%%%%%%%%%%%%%%%%%%%%%%%%%%%%%%%%%
\subsection{Cosmological Viability of the Fixed Points}
%%%%%%%%%%%%%%%%%%%%%%%%%%%%%%%%%%%%%%%%%%%%%%%%%%%%%%%%%%%%

For a model to be cosmologically viable, it must satisfy several physical conditions. For instance, the Hubble parameter must remain real, as it defines the horizon scale. In this section, we will investigate the ability of the fixed points to describe viable asymptotic cosmological scenarios.

The Hubble parameter, $H$, can be expressed in terms of the dynamical variables as: 
\begin{equation}
\label{Eq: Hubble in Variables}
    \frac{H^2}{m_\text{P}^2} = \hat{g}^2 \left(\frac{y}{z}\right)^4,
\end{equation}
where $z$ can be written in terms of $x$ and $y$ using the Friedman constraint in Eq.~\eqref{Eq: Friedman Constraint}. This formulation allows the energy scale set by $H$ to be tuned by adjusting the generalized coupling parameter $\hat{g}$, as defined in Eq.~\eqref{Eq: Generalized Charge}. This flexibility could be particularly relevant for determining the energy scale of primordial inflation or for adjusting the present-day value $H_0$ to match local observations, potentially alleviating the $H_0$ tension. 

On the other hand, since $y^4 > 0$ always holds, and $\hat{g}^2$ and $z^4$ are real and share the same sign dependence, $H^2$ is positive. This guarantees that $H$ is consistently real-valued once $z^4$ is determined at the fixed point. 

Using Eq.~\eqref{Eq: Hubble in Variables}, we can express the density and pressure of the vector field as: 
\begin{align}
    \frac{\rho_B}{m_\text{P}^4} &= 3 \hat{g}^2 \left(\frac{y}{z}\right)^4, \label{Eq: Rescaled Density} \\
    \frac{p_B}{m_\text{P}^4} &= \hat{g}^2 \left(\frac{y}{z}\right)^4 \Big[ x^2 + 2 x y + \left[1 + \left(36 c_2 - 24 c_1\right) x^2\right] y^2  \notag \\
    &+ 4 \left[\sqrt{2} (c_2 - c_1) p + 6 c_2 x\right] y^3 \nonumber \\
    &+ 4 \left[ 6 c_1 - 3 c_2 + 2 (c_2 - c_1) \epsilon \right]y^4 + 2z^4 \Big]. \label{Eq: Rescaled Pressure}
\end{align}
We note that $\rho_B > 0$ for all parameter values, as it depends on $\hat{g}^2$ and $z^4$, which share the same sign at a given fixed point. This further ensures that $H$ is a physical, real-valued quantity there.

Then, defining the re-scaled density as:
\begin{equation}
    \hat{\rho}_B \equiv \frac{\rho_B}{m_\text{P}^4 \hat{g}^2} = 3\left(\frac{y}{z}\right)^4,
\end{equation}
and evaluating it at the fixed points $A_\pm$, we find: 
\begin{equation}
    \hat{\rho}_B (A_\pm) = \frac{6}{12 c_2 - 6c_1 - \sqrt{-6c_1}},
\end{equation}
which is positive under the condition: 
\begin{equation}
\label{Eq: Condition for positive density}
    c_1 \leq 0 \quad \land \quad c_2 > \frac{c_1}{2} + \sqrt{-\frac{c_1}{24}}.
\end{equation}
Conversely, $\hat{\rho}_B < 0$ if:  
\begin{equation}
\label{Eq: Condition for negative density}
    c_1 \leq 0 \quad \land \quad c_2 < \frac{c_1}{2} + \sqrt{-\frac{c_1}{24}}.
\end{equation}
A closer inspection of the left panel in Fig.~\ref{Fig: Stability} reveals that the bifurcation curve between the attraction regions for points $A_\pm$ and $B_\pm$ is given exactly by:
\begin{equation}
    c_2 = \frac{c_1}{2} + \sqrt{-\frac{c_1}{24}},
\end{equation}
which leads to the conclusion, based on the second condition in Eq.~\eqref{Eq: Condition for negative density}, that $\hat{\rho}_B < 0$ in the attraction region of $A_\pm$ as depicted in the left panel of Fig.~\ref{Fig: Stability}. This indicates that $\hat{g}^2 < 0$, and consequently $z^4 < 0$, in this region. Conversely, $\hat{\rho}_B > 0$ outside of this attraction region, leading to $\hat{g}^2 > 0$ and $z^4 > 0$.  However, we stress that the physical energy density $\rho_B$ is positive in both regions of the parameter space. We want to clarify that effective negative energy densities are, in principle, possible in modified gravity theories, but they should not dominate the energy content, as this would lead, among other potential issues, to a negative squared Hubble parameter. For example, in a phantom dark energy model within Horndeski’s theory~\cite{Matsumoto:2017qil}, the effective energy density becomes negative only at high redshifts while exhibiting phantom behaviour at low redshifts. 

Thus, from a dynamical systems perspective, the points $A_\pm$ represent viable asymptotic states describing a de-Sitter expansion phase of the universe. When acting as attractors, $A_\pm$ could describe the late-time accelerated expansion of the universe, whereas as saddle points, they may characterize the primordial inflationary phase. To validate these interpretations, we will further investigate the dynamics of selected cosmological trajectories in phase space in a subsequent section.

%\textcolor{blue}{Therefore, if $z^4<0$ the density becomes positive and constant when the system reaches the attractor points $A_\pm$, giving rise to an effective cosmological constant at late times. However, as we will show later, the Hubble parameter has well-defined values along the curves $y = 0, \, H \to 0$ and $z = 0, \, H^2 \to \pm\infty$ (see Eq.~\eqref{Eq: Hubble in Variables}).}\\ %\\

%\textcolor{blue}{The problem with this configuration arises when the evolution transitions to $z^4 > 0$, leading to a complex Hubble parameter, or when it passes through the points $U_\pm = \{\pm1, 0\}$, where a physical singularity renders the Hubble parameter undefined (see FIG.~\ref{Fig: Evolution for the viable attractor}). Furthermore, to achieve a proper dark energy description, it is necessary for the field to allow for a matter-dominated phase before it begins to dominate. However, this requires the field to evolve in the vicinity of $U_\pm$. As we will show later, these points are unavoidable when the system evolves near them, inevitably leading to the physical singularity.
%}\\
On the other hand, at the fixed points $B_\pm$ and $C_\pm$, we find that the denominator of $\hat{\rho}_B$ vanishes, while the numerator remains constant, causing $\hat{\rho}_B$ to diverge at these points. Consequently, we conclude that these fixed points do not correspond to viable scenarios of accelerated expansion.

Summarizing:
\begin{itemize}
    \item At the points $A_\pm$, the universe undergoes an exponential accelerated expansion, which could last forever or be a transient state.
    \item At the points $B_\pm$ and $C_\pm$, $\rho_B$ goes to infinity, which makes $H$ infinite, and thus cosmologically unreliable.
\end{itemize}

Despite the fact that the fixed points $A_\pm$ represent viable cosmological solutions, it is important to note that the phase space defined by the variables in Eq.~\eqref{Eq: Dynamical System} is not compact. Specifically, both $x$ and $y$ are unbounded, meaning the system may not have a well-defined global attractor~\cite{Coley:2003mj}. This opens up the possibility for other, potentially more complex, types of asymptotic behaviour that go beyond the fixed-point analysis.

In the absence of a global attractor, the dynamics could exhibit trajectories leading to different regimes or exhibit more complex structures such as limit cycles or chaotic behaviour in certain sectors of the parameter space~\cite{Ott:2002}. To fully understand these potential outcomes, a more detailed exploration of the system’s trajectories in extended regions of the phase space is required. This could reveal additional solutions that might correspond to viable cosmological scenarios.

Next, we will delve into these possibilities by exploring the system's behaviour at the boundaries and within regions where the fixed-point analysis does not capture the full dynamical complexity.

%%%%%%%%%%%%%%%%%%%%%%%%%%%%%%%%%%%%%%%%%%%%%
\section{Pseudo Stationary States}
\label{Sec: Pseudo States}
%%%%%%%%%%%%%%%%%%%%%%%%%%%%%%%%%%%%%%%%%%%%%

%%%%%%%%%%%%%%%%%%%%%%%%%%%%%%%%%%%%%%%%%%%%%
\subsection{The Stationary Straight Lines\label{SEC: TSSL}}
%%%%%%%%%%%%%%%%%%%%%%%%%%%%%%%%%%%%%%%%%%%%%

From the preceding analysis, we conclude that only the fixed points $A_\pm$, at which accelerated expansion occurs, serve as viable asymptotic states of the universe. In contrast, at $B_\pm$ and $C_\pm$, the field density {blows up} and thus these are not viable solutions.

As discussed in Ref.~\cite{Garnica:2021fuu}, the system may admit other asymptotic fates. Specifically, it could evolve towards ``pseudo-stationary states'' at distinct scales of $x$ and $y$. Specifically in Ref.~\cite{Garnica:2021fuu}, it is shown that for large values of $x$ and $y$, the system's behaviour is governed by a linear relationship: 
\begin{equation} y = \beta x, 
\end{equation} 
where $\beta$ is a constant describing the slope of the line. In terms of the field $\psi$, this implies that the dynamics follows $\beta \dot{\psi} = H \psi$. During a de-Sitter phase, where $H$ is constant, the field enters a \textit{constant-roll regime}~\cite{Motohashi:2014ppa,Motohashi:2017vdc,Motohashi:2019tyj}, characterized by the equation: 
\begin{equation} 
\ddot{\psi} = \frac{1}{\beta} H \dot{\psi}. 
\end{equation}
In what follows, we will analyze in detail the existence and stability of this solution at large $x$.

To determine the slope of the straight line, we start by assuming $y = \beta x$. From the dynamical equation for $y$ in Eq.~\eqref{Eq: Autonomous set}, we obtain: \begin{equation} 
\label{Eq: Eq for beta} 
\frac{1}{\beta} - \frac{x'}{x} = 0. 
\end{equation} 
Next, using the equation for $x$ from Eq.~\eqref{Eq: Autonomous set}, we find: 
\begin{equation} 
\frac{x'}{x} = \epsilon + \frac{p}{\sqrt{2} x}, 
\end{equation} 
which depends only on $x$ under the assumption $y = \beta x$. In the limit $x \rightarrow \infty$, the dominant term depends solely on the constants $c_1$, $c_2$, and $\beta$. Thus, taking the limit, we obtain: 
\begin{equation} 
\lim_{x \rightarrow \infty} 
\left( \frac{1}{\beta} - \frac{x'}{x} \right) = 0, \end{equation} 
which yields the following cubic equation for $\beta$: \begin{align} 
    0 &= \left(- \frac{4}{3} + \frac{7}{3} \frac{c_2}{c_1}\right) + \left( - \frac{37}{9} + \frac{8}{9} \frac{c_1}{c_2} + \frac{56}{9} \frac{c_2}{c_1} \right) \beta \nonumber \\
    & + \left(\frac{4}{3} - \frac{2}{3} \frac{c_1}{c_2} + \frac{7}{3} \frac{c_2}{c_1} \right) \beta^2 + \beta^3.
\end{align} 
The roots of this equation are: 
\begin{align} 
    \beta_0 &= \frac{4}{3} - \frac{7}{3}\frac{c_2}{c_1}, \label{Eq: beta0} \\
    \beta_\pm &= - \frac{4}{3} + \frac{c_1}{3c_2} \pm \frac{\sqrt{(c_1 - c_2)(c_1 - 7c_2)}}{3c_2}. \label{Eq: beta+-} 
\end{align}
Therefore, for large $x$, the system evolves along a straight line defined by $y = \beta_i x$, where $\beta_i$ is one of the three slopes found in Eqs.~\eqref{Eq: beta0} and \eqref{Eq: beta+-}.

Along these lines, we find that for $y = \beta_0 x$, the equation of state parameter is $w_B = -1$, corresponding to a de-Sitter phase. For the cases where $y = \beta_\pm x$, we find: 
\begin{equation} 
w_B = -\frac{23}{9} + \frac{8}{9} \frac{c_1}{c_2} \pm \frac{8}{9c_2} \sqrt{(c_1 - c_2)(c_1 - 7c_2)}. \label{Eq: Equation of state beta}\end{equation} 
Although all these three lines may correspond to accelerated solutions, only the first one describes a de-Sitter phase. Thus, to keep our presentation simple, we will focus on the system's behaviour around this line in the following sections and leave the discussion of the other lines for Appendix~\ref{App: Straight Lines}.

As noted in Ref.~\cite{Garnica:2021fuu}, if $\beta_0$ is negative, the system evolves toward smaller values of $x$ and $y$, and as $y$ approaches zero, the system becomes dominated by the lower powers of the field, represented by the Yang-Mills Lagrangian. This leads to an exit from the accelerated phase into a decelerated expansion, where the vector field behaves like a radiation fluid. Conversely, a positive slope describes a system moving toward larger values of $x$ and $y$, potentially describing a phase of dark energy domination~\cite{Rodriguez:2017wkg}. However, this potential dark energy domination was not explored in sufficient detail in Ref.~\cite{Rodriguez:2017wkg}. Therefore, we will investigate this scenario further here. As a result, we will demonstrate that the inflationary phase, either primordial or late-time, exhibits several shortcomings leading to nonphysical outcomes, thereby challenging the viability of the GSU2P theory as a cosmological model.

%%%%%%%%%%%%%%%%%%%%%%%%%%%%%%%%%%%%%%%%%%%%%
\subsection{Existence of the Central Zone}
%%%%%%%%%%%%%%%%%%%%%%%%%%%%%%%%%%%%%%%%%%%%%

After analyzing the behaviour of the system in the large-value regime of the variables $x$ and $y$, we now turn our attention to the opposite regime in which these variables are small. This regime is cosmologically relevant for two main reasons. First, in a primordial inflationary model, the field driving the accelerated expansion is expected to decay near the end of inflation, giving place to the reheating process. As a result, while trajectories may begin in the large-value regime of $x$ and $y$, they are expected to evolve towards smaller values of these variables as inflation concludes. Second, to accurately reproduce the post-Big Bang expansion history of the universe, the dark energy-dominated epoch must be preceded by a phase of decelerated expansion dominated by pressureless matter. During this matter-dominated phase, the vector field should be subdominant, leading to trajectories in the phase space evolving with $y$ close to zero.

As pointed out in Ref.~\cite{Garnica:2021fuu}, small values of $y$ can lead to singularities in the system, as the denominator of $x'$ in Eq.~\eqref{Eq: Autonomous set} approaches zero, causing the system to diverge. This denominator is expressed as:
\begin{align}
\label{Eq: Denominator fx}
    D_{x'} &= 16 (c_1 - 7 c_2) (c_1 - c_2) y^7 + 8 (c_1 - c_2) y^5  \nonumber \\
    &- 12 c_2 y^3 + y.
\end{align} 
This singularity can occur in a region referred to as ``central zone'' of the phase space in which the system can enter provided that $D_{x'}$ does not vanish. However, the conditions under which this central zone arises, as well as its implications for the dynamics of the system, remain unexplored. In the subsequent analysis, we will investigate this issue in detail.

In non-compact phase spaces, the study of \textit{nullclines}—the geometric curves where $x_i' = 0$, for a given variable independent of the others—can reveal important classes of asymptotic behaviour of the system. Although nullclines do not correspond to true fixed points, they can provide insight into the existence of ``pseudo-stationary states'' within the system~\cite{Ott:2002}. We will show that the existence of the central zone is guaranteed by two of such pseudo-stationary states, which arise from the nullcline of the variable $x$. 

Solving the equation $x' = 0$, when $y \rightarrow 0$, we find the points:
\begin{equation}
    U_\pm = \{\pm 1, 0\}.
\end{equation}
To analyze the stability of $U_\pm$, we compute the eigenvalues, $\lambda_i^\pm$, and eigenvectors, $\nu_i^\pm$, of the Jacobian matrix evaluated at $x = \pm 1$,  obtaining the dominant terms for small $y$. For the point $U_+$, we find:
\begin{align}
U_+&: \quad \lambda_1^+ = - \frac{1}{2} y, \quad \lambda_2^+ = 3 + \frac{4}{y}, \nonumber \\
&\nu_1^+ = \{ 0, 1 \}, \quad \nu_2^+ = \left\{3 + \frac{4}{y}, 1 \right\},
\end{align}
whereas for $U_-$, the corresponding expressions are:
\begin{align}
U_-&: \quad \lambda_1^- = + \frac{1}{2} y, \quad \lambda_2^- = 3 - \frac{4}{y}, \nonumber \\
&\nu_1^- = \{ 0, 1 \}, \quad \nu_2^- = \left\{3 - \frac{4}{y}, 1 \right\}.
\end{align}
These results indicate that $U_\pm$ are saddle-like points. A saddle point is characterized by some trajectories in phase space moving towards the fixed point while others moving away from it.

Notice that $\nu_1^\pm$ are unitary vectors pointing towards the $y$-direction, whereas the $x$-component  of the eigenvectors $\nu_2^\pm$ depends on the value of $y$. Notably, these $x$-components become large as $y$ approaches zero. For $U_\pm$, we must consider two cases: $y \rightarrow 0^+$ and $y \rightarrow 0^-$. When $y \rightarrow 0^+$, we have that $\lambda_2^-$ is negative and $\nu_2^- \approx \{-\infty, 1\}$, while $\lambda_2^+$ is positive and $\nu_2^+ \approx \{\infty, 1\}$. As a result, when $y$ takes on positive values, trajectories in phase space around $U_-$ converge towards it, while those near $U_+$ diverge from it. Conversely, when $y \rightarrow 0^-$, $\lambda_2^-$ becomes positive, and $\nu_2^- \approx \{\infty, 1\}$, causing trajectories to move away from $U_-$; simultaneously, $\lambda_2^+$ turns negative, and $\nu_2^+ \approx \{-\infty, 1\}$, leading trajectories to converge towards $U_+$. In both cases, the remaining eigenvalue—$\lambda_1^+$ for $U_+$ and $\lambda_1^-$ for $U_-$—changes sign, confirming the saddle-like nature of these pseudo-critical points.

\begin{figure}[t!]
\centering    
\includegraphics[width=0.48\textwidth]{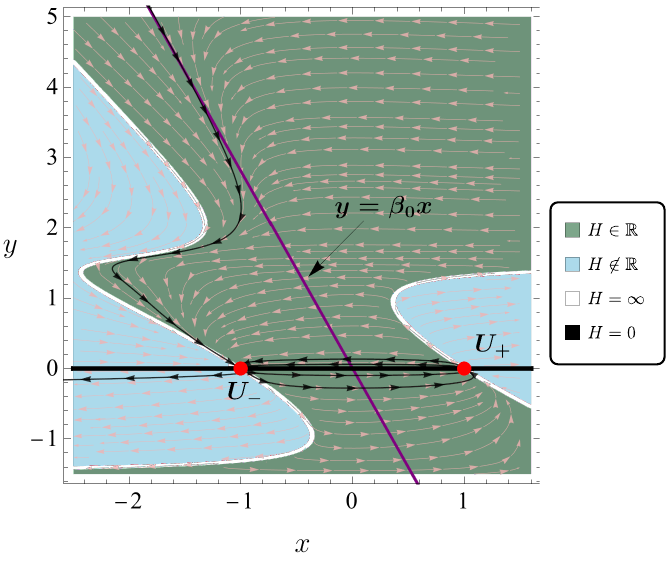}
\caption{Phase space evolution of a specific trajectory (line with arrows) using initial conditions and parameters from Fig. 1(a) in Ref.~\cite{Garnica:2021fuu}: $x_i = 5 \times 10^9$, $y_i = 10^{10}$, $\alpha_1 = 1$, $\alpha_3 = 1.0008$, and $\chi_5 = -1.965$, corresponding to $c_1 = 0.0206$ and $c_2 = 0.0366$. The trajectory initially follows the attractor line $y = \beta_0 x$ (purple line), where $z^4 > 0$, fixing $\hat{g}^2 > 0$, and the Hubble parameter takes on real values (green region). As the system evolves, $z^4 \rightarrow 0$ and thus $H \rightarrow \pm \infty$ (white lines). Eventually, the trajectory enters the central zone, and oscillates between the points $U_\pm$ (red dots), which lies on a line where $H = 0$. Then, after some oscillations, it finally escapes from the central zone due to the saddle instability of $U_-$. When exiting from the central zone, $z^4$ flips sign causing the Hubble parameter to become complex (light blue region).}
\label{Fig: Example Phase Space}
\end{figure}

We conclude that when $y \rightarrow 0^+$,  trajectories in phase space approach $U_-$ causing $y$ to become negative and move towards $U_+$. Upon reaching $U_+$, $y$ becomes positive again and is subsequently attracted back to $U_-$. This oscillating behaviour, characterized by alternating repulsion and attraction between the points $U_\pm$ generates what we call the ``central zone''. This feature will be evidenced numerically.

\begin{figure*}[t!]
\centering    
{\includegraphics[width=0.48\textwidth]{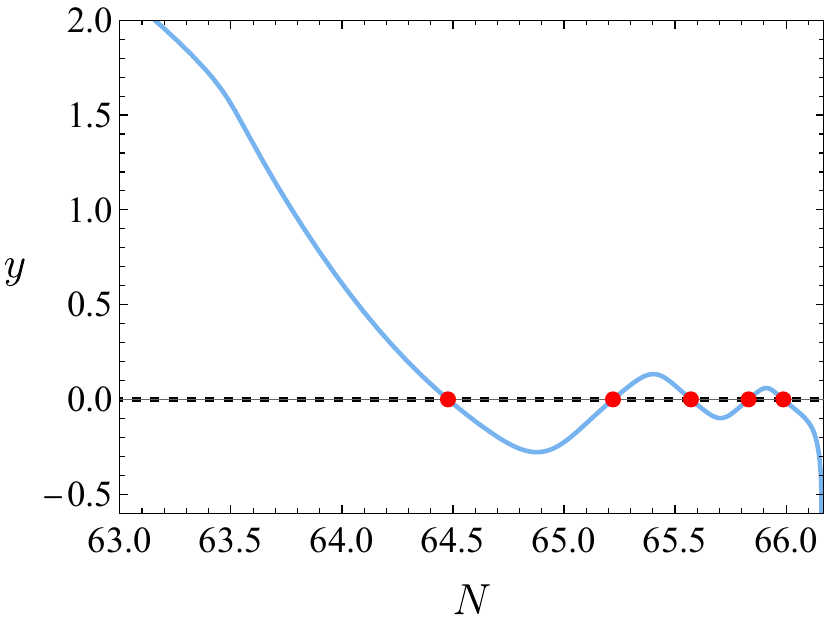}} \hfill
{\includegraphics[width=0.48\textwidth]{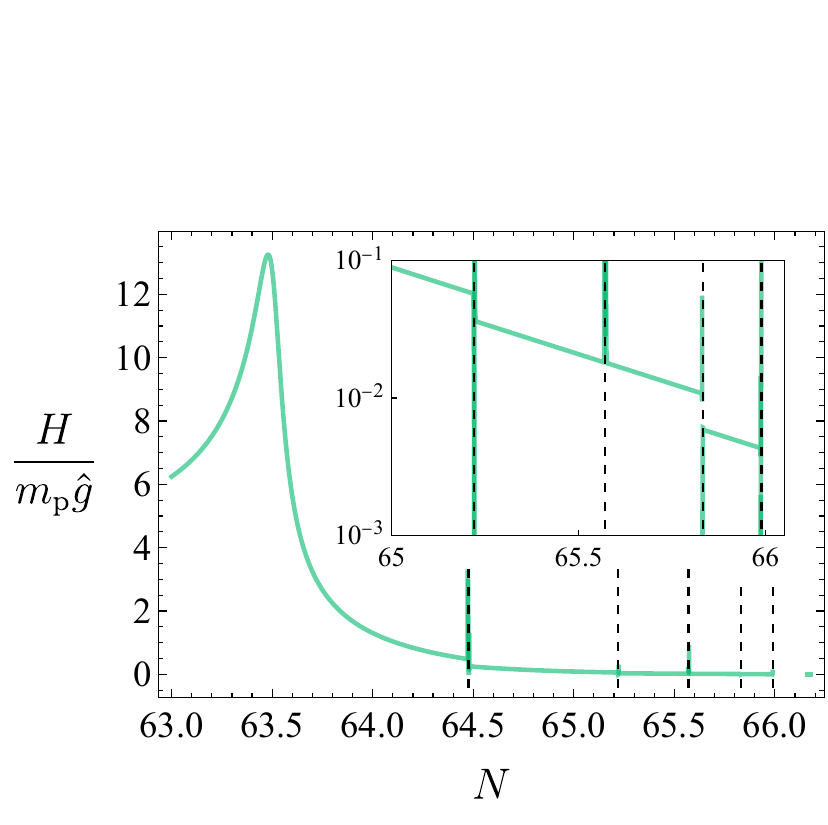}}
\caption{(Left) Time evolution of the variable $y$, with the red dots indicating the precise moments at which $y$ crosses cero, the first one occurring around $N = 64.5$. When $y$ crosses zero, 
%$\hat{\rho}_B$ flips its sign as described by Eq.~\eqref{Eq: Approximated density}.
{$H$ becomes indeterminate.}
(Right) Time evolution of the scaled Hubble parameter $H/(m_\text{p}\hat{g})$. At each crossing, $H$ exhibits undefined values, marked by the thin black dashed lines, corresponding to the physical singularities within the central zone.} 

%Time evolution of the rescaled density $\hat{\rho}_B$. {When $y$ crosses zero the first time}, $\hat{\rho}_B$ exhibits rapid (but small) oscillations, alternating between negative and positive values, reflecting the system's behaviour within the central zone.}
\label{Fig: Instability}
\end{figure*}

%%%%%%%%%%%%%%%%%%%%%%%%%%%%%%%%%%%%%%%%%%%%%%%%%%
\subsection{Issues in the Central Zone}
%%%%%%%%%%%%%%%%%%%%%%%%%%%%%%%%%%%%%%%%%%%%%%%%%%

As previously discussed, the bouncing behaviour between the points $U_\pm$ gives rise to what we call the central zone. However, it is important to emphasize that these points, $U_\pm$, are saddle-like pseudo-stationary states which arise from the nullcline for the variable $x$ when $y \rightarrow 0$.  Consequently, the bouncing behaviour does not necessarily lead to a limit cycle—self-sustained oscillations around a point, where the system follows a closed trajectory that repeats periodically, and any small perturbation causes the system to return to this trajectory~\cite{Ott:2002}. In the following, we will delve into these issues by numerically investigating the system's dynamics around the central zone. %and demonstrate the non-existence of such a limit cycle in the central zone, thus confirming the system's instability in this region.

%%%%%%%%%%%%%%%%%%%%%%%%%%%%%%%%%%%%%%%%%%%%%%%%%%%%%%%%%%%%%%%%%
\subsubsection{\textbf{Issues for the Inflationary Scenario}}
%%%%%%%%%%%%%%%%%%%%%%%%%%%%%%%%%%%%%%%%%%%%%%%%%%%%%%%%%%%%%%%%%

To illustrate the system's dynamics within the central zone, we consider the phase space $\{x, y\}$ for a specific parameter set $\{c_1, c_2\}$. We focus on the trajectory represented by the line with arrows depicted in Fig.~\ref{Fig: Example Phase Space}. This trajectory shows the system's evolution starting with initial conditions drawn from Fig. 1(a) of Ref.~\cite{Garnica:2021fuu}:
\begin{equation}
\label{Eq: ICs Fig. 1(a)}
    x_i = 5 \times 10^9, \quad y_i = 10^{10},
\end{equation}
with parameter values:
\begin{equation}
\alpha_1 = 1, \quad \alpha_3 = 1.0008, \quad \chi_5 = -1.965.
\end{equation}
Using the relations from Eq.~\eqref{Eq: New Constants}, these parameters are translated into our system parameters as:
\begin{equation}
    c_1 = 0.0206, \quad c_2 = 0.0366.
\end{equation}
For these parameters, none of the fixed points $A_\pm$, $B_\pm$, or $C_\pm$ are real, leaving only the pseudo-stationary points $U_\pm$ visible. The system's trajectory initially follows the attractor line $y = \beta_0 x$ (depicted by the solid purple line) towards decreasing values of $y$. During this phase, $z^4 > 0$, fixing the sign of $\hat{g}^2$ as positive, and the Hubble parameter takes on real values (green region) [see Eq.~\eqref{Eq: Hubble in Variables}]. As $y$ decreases from positive values, the trajectory approaches $U_-$, $z^4 \rightarrow 0$, and thus $H$ becomes indeterminate (white line). Note that the points $U_\pm$ are located at the intersection of the white lines (where $H \rightarrow \infty$) with the black line (where $H = 0$). Therefore, at each crossing through $U_\pm$, the expansion rate becomes undefined, highlighting a critical dynamical issue in the model.

As noted in Ref.~\cite{Garnica:2021fuu}, entering the central zone typically leads to oscillatory behaviour, marking the end of the primordial inflationary phase. We argue that this access inherently introduces a dynamical inconsistency into the system. For trajectories approaching the point $U_-$ with $x \rightarrow - 1$ and $y \rightarrow 0$, the density and pressure of the vector field, according to Eqs.~\eqref{Eq: Rescaled Density} and \eqref{Eq: Rescaled Pressure}, can be approximated as:
\begin{equation}
\label{Eq: Approximated density}
    \frac{\rho_B}{\hat{g}^2 m_\text{P}^4} \approx 3 y^3, \quad \frac{p_B}{\hat{g}^2 m_\text{P}^4} \approx y^3,
\end{equation}
indicating that the system behaves like a radiation fluid with $w_B \approx 1/3$ around this point. Then, upon reaching $U_-$, $y$ becomes negative, causing the trajectory to escape from $U_-$ and move towards $U_+$ as $y$ approaches 0 from the negative side. Once again, when the trajectory reaches $U_+$, the trajectory is repelled as $y$ becomes positive and subsequently it is attracted towards $U_-$ again since $y \rightarrow 0^+$. The system oscillates between $U_-$ and $U_+$ with $H \in \mathbb{R}$, but $H$ becomes indeterminate at each crossing.\footnote{It is worth clarifying that the approximation presented in Eq.~\eqref{Eq: Approximated density} is valid in the vicinity of $U_-$ but just before the trajectory originating outside the central zone reaches it. Inside the central region, i.e., after the above mentioned trajectory reaches $U_-$, the behaviour of the density and pressure  are of the form $\rho_B \propto 6y^4$ and $p_B \propto 2y^4$ respectively, the proportionality factor being the same for both quantities.  The system, then, neither exhibits negative energy density nor negative pressure.}

This dynamics continue until the saddle point nature of $U_-$ eventually forces the trajectory out of the central zone, escaping towards infinity in the phase space, as no attractors exist in that region. When the trajectory escapes from the central zone, $z^4$ flips sign and thus $H$ becomes complex (blue region). This transition from real to complex values in $H$ signals a flaw in the theory.

Notably, this shortcoming arises near the end of the primordial inflationary phase. Following the constant-roll condition $y = \beta_0 x$, the amount of inflation can be calculated as:
\begin{equation}
\label{Eq: Expected N_inf}
    N_\text{inf} \equiv \int \text{d} t \ H = \int_{y_i}^{y_f} \beta_0 \frac{\text{d}y}{y} \approx - \beta_0 \ln y_i,
\end{equation}
where the magnitude of $y$ at the end of inflation, denoted as $y_f$, is neglected in comparison with its initial value $y_i$. Numerically, this translates into:
\begin{equation}
    N_\text{inf} \approx 64.75.
\end{equation}
For the initial conditions in Eq.~\eqref{Eq: ICs Fig. 1(a)}, this issue is numerically confirmed in Fig.~\ref{Fig: Instability}. The left panel shows the evolution of $y$ over time, with red dots marking the moments when $y$ crosses zero—corresponding to the system passing through the points $U_\pm$. The first crossing occurs approximately at $N = 64.5$. The right panel depicts the corresponding evolution of the rescaled expansion rate in Eq.~\eqref{Eq: Hubble in Variables}, which becomes indeterminate precisely at these crossings.
 
\begin{figure}[t!]
\centering    
\includegraphics[width=0.45\textwidth]{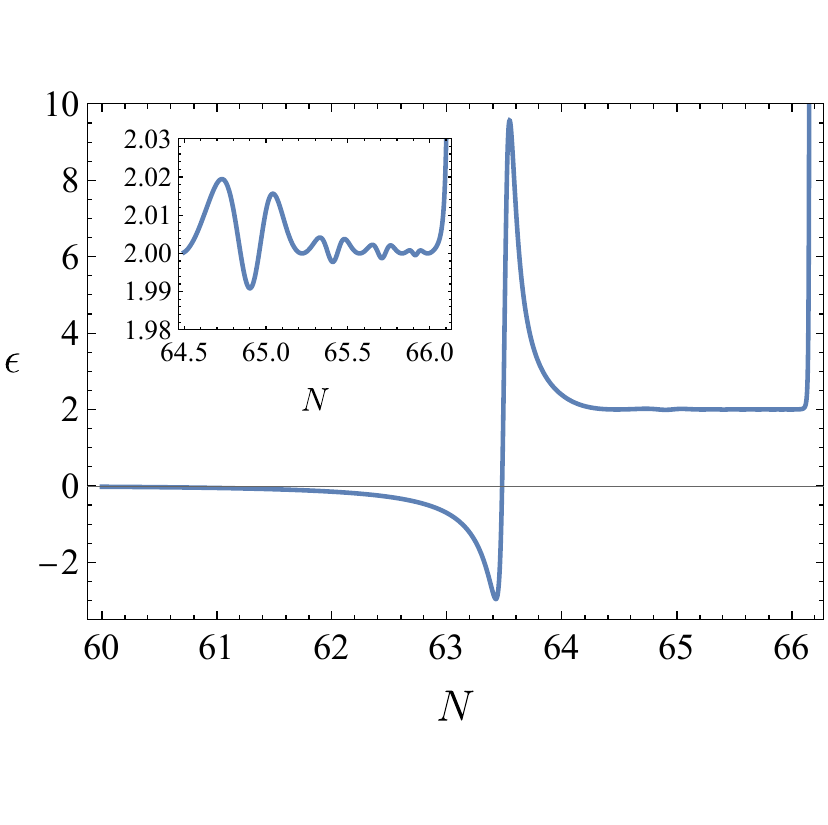}
\caption{Evolution of the parameter $\epsilon$. Initially, $\epsilon \approx 0$, indicating that the system is in a constant-roll phase. Upon entering the central zone, $\epsilon$ exhibits oscillations around $\epsilon \approx 2$, reflecting the system's behaviour as a radiation fluid. Finally, the system exits the central zone, resulting in an uncontrolled growth of the field. As highlighted in the inset, this growth ultimately leads to the divergence of $\epsilon$.
%Two irregularities are evident in the evolution of $\epsilon$. The first occurs as the vector field's density flips sign upon entering the central zone. The second arises when the system exits the central zone, causing the field to grow uncontrollably. As highlighted in the inset, this uncontrolled growth ultimately results in a divergence of $\epsilon$.
}
\label{Fig: Instability epsilon}
\end{figure}

The system exhibits another significant issue, stemming from the absence of a limit cycle in the central zone. As shown in Fig. \ref{Fig: Example Phase Space}, the trajectory oscillates between the points $U_\pm$ until, due to the saddle-like nature of $U_-$, eventually forces it to escape the central zone. Upon exiting this region, the lack of an attractor leads to an unbounded growth in the field magnitude and its velocity after approximately 66 $e$-folds, further highlighting the system's indeterminacy. 
{This behaviour is directly reflected in the parameter $\epsilon$, as illustrated in Fig.~\ref{Fig: Instability epsilon}.} Initially, $\epsilon \approx 0$ indicates that the system is in a constant-roll phase. Upon entering the central zone, $\epsilon \approx 2$, which reflects the system's behaviour as a radiation fluid. A natural interpretation—{which is biased} by the indeterminations in $H$ shown in the right-hand side of Fig.~\ref{Fig: Instability}—is to associate the transition from the accelerated phase to the decelerated one (corresponding to the first spike in $\epsilon$ in Fig.~\ref{Fig: Instability epsilon}) to some singularity in $H$. However, as shown by the behaviour of $H$ (right-hand side of Fig.~\ref{Fig: Instability}), the Hubble parameter remains well behaved during this transition, which begins at $N=63.5$ and ends near $N=64.5$. 

Furthermore, choosing $\epsilon$ as the main observable {might conceal} physical singularities because it behaves smoothly during the transition from the accelerated phase to the decelerated one. Afterwards, $\epsilon$ exhibits the typical oscillatory behaviour present in all viable models of inflation, without any of the singularities observed in $H$ in Fig.~\ref{Fig: Instability}. This behaviour suggests a graceful exit from the inflationary phase into the radiation-dominated era. However, this interpretation is delicate, as $H$ becomes indeterminate once the system reaches the $U_\pm$ points.  At this stage, the system may either escape the central zone or remain within it, depending on the tuning of the initial conditions and the integration interval. If the system exits the central zone, it experiences uncontrolled growth in $\epsilon$, ultimately leading to a divergence, as illustrated at the end of the inset plot of Fig.~\ref{Fig: Instability epsilon}.

%This behaviour is also reflected in the parameter $\epsilon$, as illustrated in Fig.~\ref{Fig: Instability epsilon}. Initially, $\epsilon \approx 0$ indicates the system is in a constant-roll phase. Upon entering the central zone, $\epsilon \approx 2$, reflecting the system's behaviour as a radiation fluid. However, two singularities emerge. The first occurs during the transition from the accelerated to the decelerated phase, where a sign flip in the density near $N = 64.5$ causes an indeterminacy in $H$, evident as a rapid spike in $\epsilon$. The inset plot highlights the second issue: upon exiting the central zone, the system experiences uncontrolled growth in $\epsilon$, ultimately resulting in a divergence.

In summary, trajectories that begin in the large-value regime of the variables $x$ and $y$ initially follow the attractor line $y = \beta_0 x$ until they approach the central zone. Upon entering this zone, the trajectories start oscillating between the pseudo-stationary states $U_\pm$, with the field behaving like a radiation fluid. However, during each crossing, the expansion rate becomes undefined, revealing a significant flaw in the model. After a limited number of oscillations, the trajectories inevitably exit the central zone, resulting in uncontrolled growth and divergence in the system. Moreover, as the trajectories escape, $z^4$ reverses sign, causing $H$ to acquire complex values.

%%%%%%%%%%%%%%%%%%%%%%%%%%%%%%%%%%%%%%%%%%%%%%%%%%%%%%%%%%%%%%%%
\subsubsection{\textbf{Issues for the Dark Energy Scenario}}
%%%%%%%%%%%%%%%%%%%%%%%%%%%%%%%%%%%%%%%%%%%%%%%%%%%%%%%%%%%%%%%%

%\begin{figure}
%    \centering
%    \includegraphics[width=0.48\textwidth]{AttractorASingulariy.pdf}
%\caption{Phase space evolution of a specific trajectory with initial conditions $x_i = -1.5$, $y_i = 0.5$ and parameters $c_1 = -0.784$ and $c_2 = -2.99$. The trajectory initially follows the curve $z = 0$, enters the central region, and oscillates between the points $U_\pm = \{\pm1, 0\}$. Eventually, it escapes the central region due to the saddle point instability of $U_+$ and reaches the attractor point $A_+$. The blue region ($z^4>0$) indicates where the Hubble parameter takes complex values, while the white line ($z=0$) corresponds to $H^2 = \pm\infty$. The black straight line ($y=0$) represents $H = 0$, and in the green region ($z^4<0$), the Hubble parameter is real, indicating physical values.}
%\label{Fig: Evolution for the viable attractor}
%\end{figure}

\begin{figure*}[t!]
\centering    
{\includegraphics[width=0.43\textwidth]{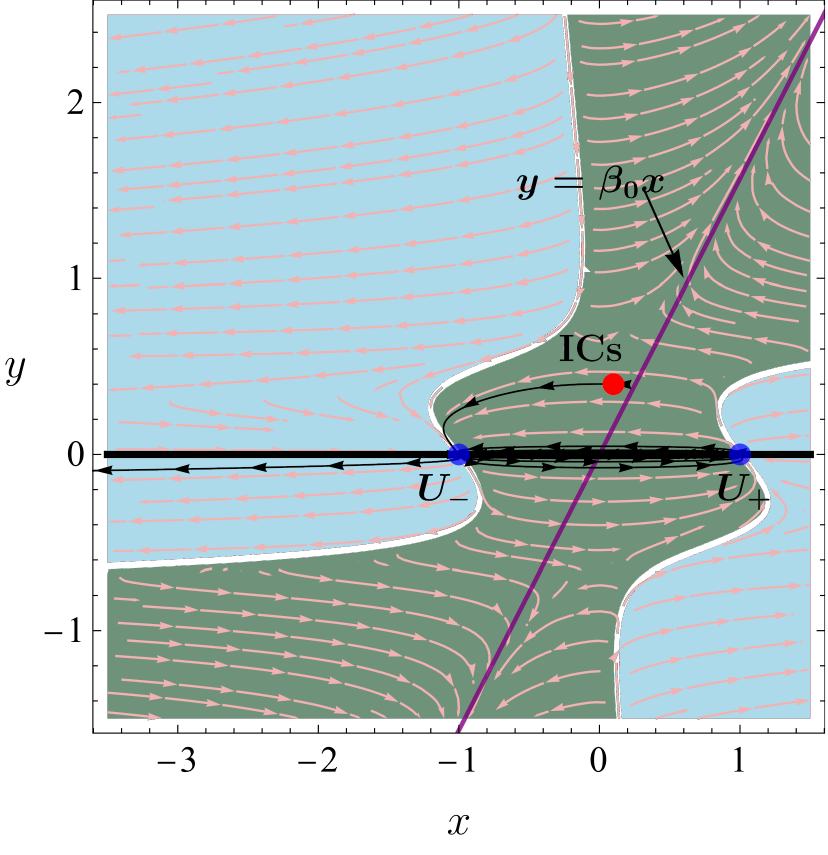}} \hfill
{\includegraphics[width=0.53\textwidth]{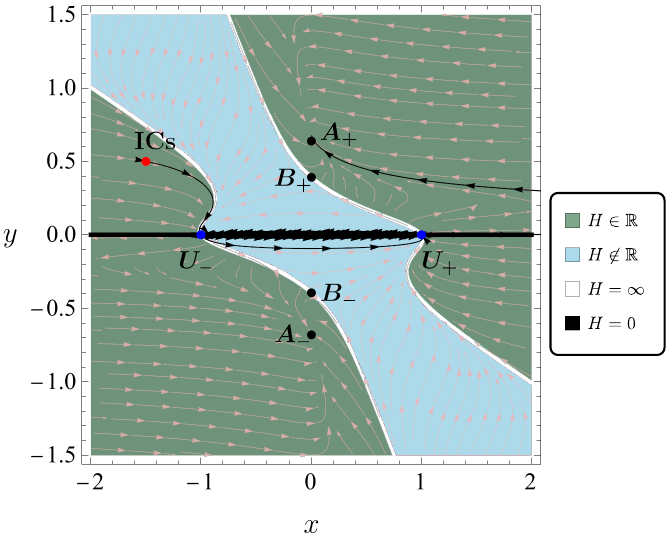}}
\caption{(Left) Evolution of a trajectory in phase space (line with arrows) starting within the central zone. The parameters are $c_1 = -1$ and $c_2 = 0.1$, with initial conditions $x_i = 0.1$ and $y_i = 0.4$ (red dot). After a few oscillations, the system escapes the central zone and diverges as the trajectory fails to align with the attractor line (purple line) with a positive slope. (Right) Phase space evolution of a specific trajectory with initial conditions $x_i = -1.5$, $y_i = 0.5$ and parameters $c_1 = -0.784$ and $c_2 = -2.99$, ensuring $A_+$ is an attractor. The trajectory initially follows the curve $z = 0$, enters the central region, and oscillates between the points $U_\pm$. Eventually, it escapes the central region due to the saddle point instability of $U_+$ and reaches the attractor point $A_+$. Although the evolution starts with a real-valued $H$ (green region), it becomes complex within the central zone due to a flip sign of $z^4$ (light blue region). Finally, $H$ becomes real once more when the trajectory escapes from the central zone to approach the attractor $A_+$. Moreover, at each crossing through the points $U_\pm$, $H$ becomes indeterminate. This behaviour of the system forbids the transition from a matter dominated epoch to a dark energy dominated epoch, and thus, the GSU2P theory is not suitable to describe the cosmic history of the universe.}
%The blue region ($z^4>0$) indicates where the Hubble parameter takes complex values, while the white line ($z=0$) corresponds to $H^2 = \pm\infty$. The black straight line ($y=0$) represents $H = 0$, and in the green region ($z^4<0$), the Hubble parameter is real, indicating physical values.}
\label{Fig: Central Zone}
\end{figure*}

For the GSU2P theory to be a plausible candidate for describing the late-time accelerated expansion of the universe, the field must remain subdominant during earlier stages of cosmic history. Specifically, the domination of the vector field driving the accelerated expansion should be preceded by a matter-dominated epoch, ensuring that $\rho_B \ll \rho_m$ during that phase. This requires that trajectories in phase space originate within the central zone, where $x$ and $y$ are small. From there, the system must follow the straight line $y = \beta_0 x$ (with $\beta_0 > 0$) allowing the field to adhere to the constant-roll dynamics, or alternatively, approach the attractor points $A_\pm$.

In the case when there is no attractor point, when a trajectory escapes from the central zone, it diverges, as illustrated in the left panel of Fig.~\ref{Fig: Central Zone}. This plot uses the parameters $c_1 = -1$, and $c_2 = 0.1$, such that $\beta_0 > 0$, with initial conditions set within the central zone: $x_i = -0.15$ and $y_i = 0.6$. The trajectory (line with arrows) starts from the initial conditions (marked by a red dot) fixing the expansion rate as real (green region), moves toward the pseudo-stationary state $U_-$, oscillates a few times between $U_\pm$ (blue points), where $H$ becomes indeterminate at each crossing (intersection between the white lines and the black line), and then escapes from the central zone through $U_-$ to a region where there is no any attractor point, leading to divergence since the trajectory is not able to approach to the attractor line with positive slope (purple line). Moreover, after escaping the central zone, the expansion rate becomes complex (light blue region). Numerical integration of the autonomous set shows that these oscillations take around $1.24$ $e$-folds in total, after which the system becomes non-integrable due to the divergence.

From this numerical example, we conclude that although the field begins as a subdominant component of the cosmic budget (subdominant to the matter fluid), it rapidly becomes dominant after only a few $e$-folds.\footnote{As a reference, the time between photon decoupling and the present day corresponds to around 7 $e$-folds. Additionally, the radiation-dominated epoch is expected to last at least 15 $e$-folds~\cite{Alvarez:2019ue}.} Once the field transitions from its radiation-like behaviour, the system diverges, making it impossible to achieve a viable cosmological scenario in which an early radiation-dominated epoch is followed by matter domination and eventually leads to an era dominated by the vector field driving the accelerated expansion of the universe.

In the case $A_\pm$ serve as attractors, the system does not diverge but other issues arise. For the right panel of Fig.~\ref{Fig: Central Zone}, we use the parameters $c_1 = -0.784$, and $c_2 = -2.99$, ensuring that $A_\pm$ are attractors. The initial conditions are chosen outside the central zone as $x_i = -1.5$ and $y_i = 0.4$ (red dot). The trajectory begins at these initial conditions, rapidly moves toward the pseudo-stationary state $U_-$, oscillates a few times between $U_\pm$ (blue points), and then escapes from the central zone through $U_+$, leading to the attractor $A_+$ (black point). However, the system exhibits a significant issue: while the trajectory starts with a real-valued expansion rate $H$ (green region) as it approaches $U_-$, $H$ becomes indeterminate (at the intersection of the black and white lines). Within the central zone, the Hubble parameter is no longer real. Finally, when the trajectory escapes from the central zone and reaches the attractor $A_+$, $H$ becomes real again. The entire process, from the initial conditions to reaching the attractor point $A_+$, spans approximately 4 $e$-folds, which is not long enough to cover the periods of radiation and matter domination. Therefore, although the attractor points $A_\pm$ represent viable cosmological solutions, the system cannot reach them from within the central zone, since $H$ is not real there. This implies that the system is not physically allowed to evolve within the central zone, which is crucial to ensure that dark energy remains subdominant before reaching its attractor, where it would drive perpetual cosmic acceleration. An alternative approach is either to choose initial conditions close to the attractor points $A_\pm$ or to adjust the parameters to avoid a divergence. However, this strategy poses significant challenges. If the initial conditions are set such that $y$ is close to $0$ and $x$ is large, the stability of the points $U_\pm$ remains independent of the parameters; this inevitably leads to uncontrolled growth of the field, causing it to dominate rapidly (within less than $3$ $e$-folds) over matter or radiation. Alternatively, selecting initial conditions after the uncontrolled growth has occurred allows the system to reach the attractor quickly without encountering a singularity; however, this requires fine-tuning the parameter $\hat{g}$ to slow the field's domination and allow matter and radiation to persist. In this case, the field rapidly stabilizes at the attractor, effectively behaving as a constant, which is indistinguishable from the $\Lambda$CDM model.

%%%%%%%%%%%%%%%%%%%%%%%%%%%%%%%%%%%%%%%%%%%%%%%%%%%%%%
\subsection{Regularization of the Autonomous Set}
%%%%%%%%%%%%%%%%%%%%%%%%%%%%%%%%%%%%%%%%%%%%%%%%%%%%%%

\begin{figure}[t!]
\centering    
\includegraphics[width=0.45\textwidth]{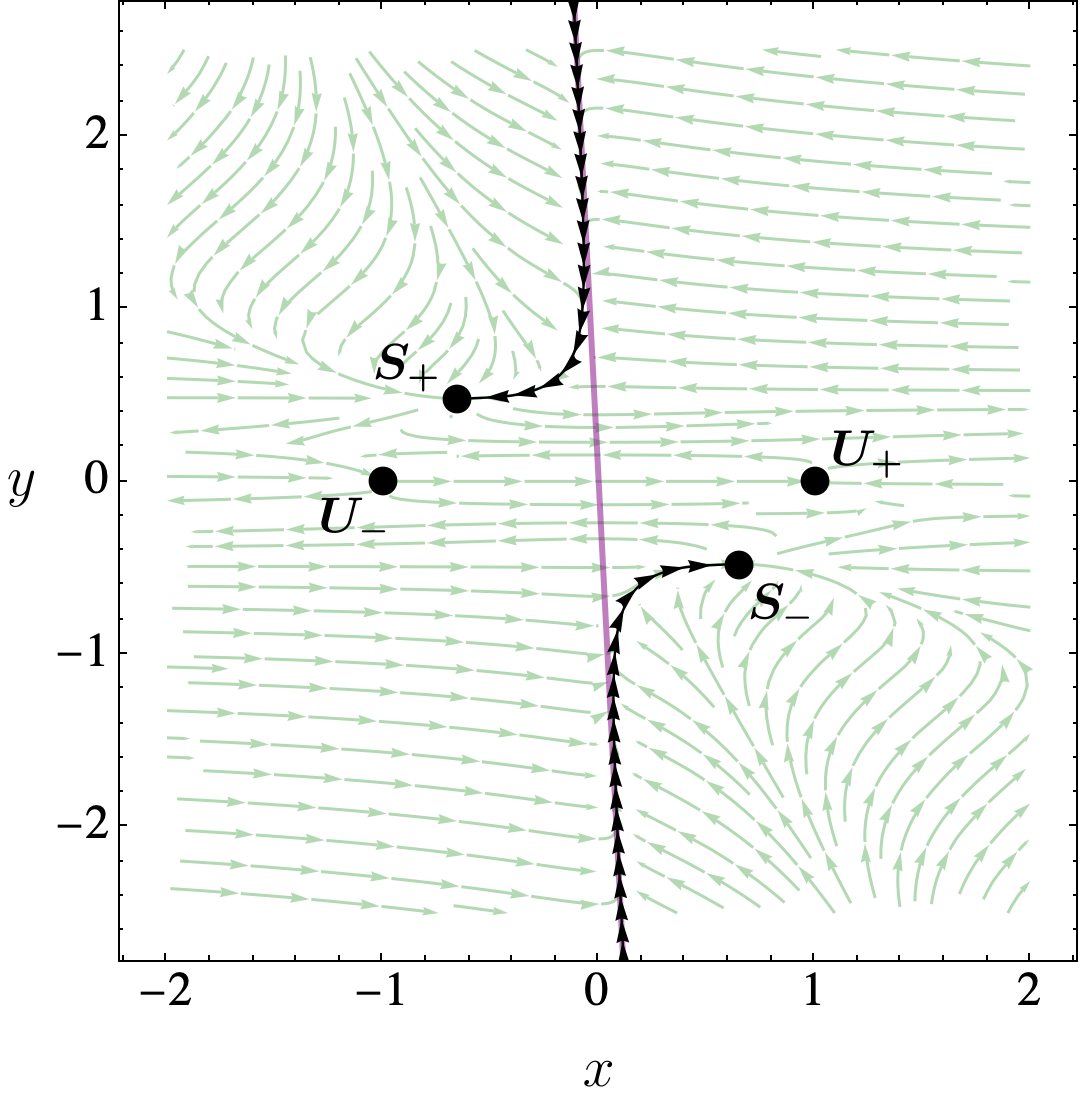}
\caption{Evolution of two trajectories in phase space with initial conditions $x_i = \pm 4 \times 10^9$ and $y_i = \mp 10^{10}$, using parameters $c_1 = 0.2$ and $c_2 = 2.2$. The trajectories initially follow the attractor line $y = \beta_0 x$, but eventually approach the singularity points $S_\pm$, determined by the regularization of the dynamical equation for $x$. After approximately 553.4 e-folds, the trajectories reach these singularities, at which point the system becomes non-integrable, causing the dynamics to cease.}
\label{Fig: No Central Zone}
\end{figure}

\begin{figure*}[t!]
\centering    
{\includegraphics[width=0.46\textwidth]{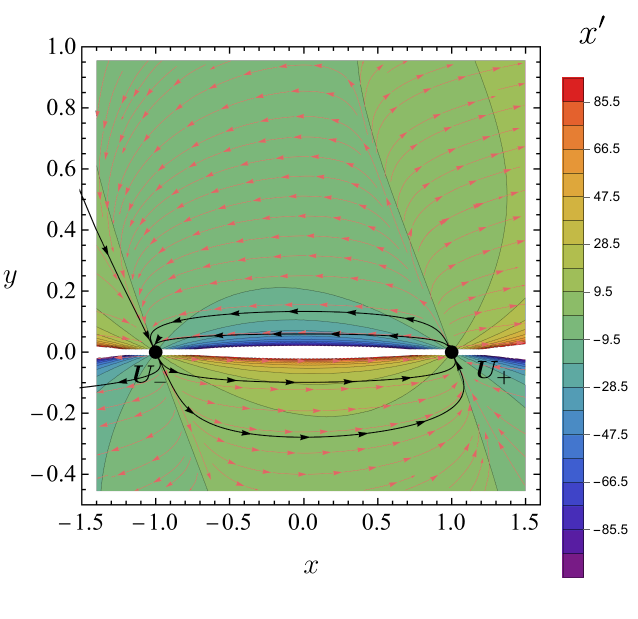}} \hfill
{\includegraphics[width=0.46\textwidth]{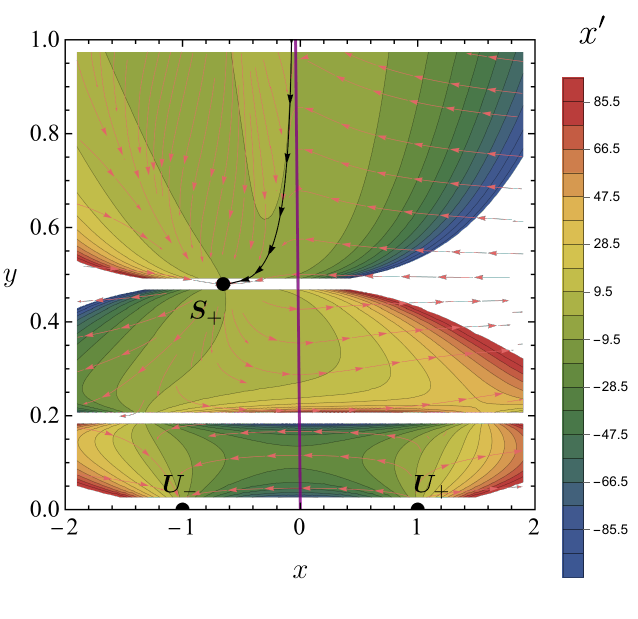}}
\caption{(Left) Evolution of a phase space trajectory using the same parameters and initial conditions as in Fig.~\ref{Fig: Example Phase Space}. After a few oscillations, the trajectory exits the central zone, reaching the white zone and finally diverging. However, the regularization points $U_\pm$ can be bypassed during the integration process since the white region width is smaller enough than the numerical integration step. (Right) Evolution of a phase space trajectory with the same parameters and initial conditions as in Fig.~\ref{Fig: No Central Zone}. In this case, the regularization point $S_+$ is surrounded by a larger white region (where $x'$ is large), rendering numerical integration infeasible.}
\label{Fig: Numerical singularity}
\end{figure*}

In the previous sections, we have analyzed the system's dynamics both outside the central zone, following the attractor line $y = \beta_0 x$ (where constant-roll occurs), and within the central zone when initial conditions are close to zero. We have demonstrated that the system is indeterminate in both cases, as it eventually escapes the central zone after oscillating between the saddle-like pseudo-stationary states $U_\pm$ for a few $e$-folds. However, due to the non-compact nature of the phase space, it is also possible to encounter singularities caused by the presence of incomplete nullcline curves, i.e., points where a given dynamical equation becomes indeterminate. In what follows, we will investigate the existence of such singularities in phase space.

As noted in Ref.~\cite{Garnica:2021fuu}, the system might not enter the central zone when approaching from the attractor line if the parameters are not properly chosen, leading to the vanishing of the denominator in the dynamical equation for the variable $x$ [see Eq.~\eqref{Eq: Denominator fx}]. The expression for this denominator, $D_{x'}$, can be written as a seventh-degree polynomial:
\begin{equation}
    D_{x'} = y(1 - \alpha y^2 + \gamma y^4 + \delta y^6),
    \label{Eq: Denominator}
\end{equation}
where
\begin{equation}
    \alpha \equiv 12c_2, \quad \gamma \equiv 8(c_1 - c_2), \quad \delta \equiv 2\gamma(c_1 - 7c_2).
\end{equation}
Since the expression inside the parentheses in Eq. (\ref{Eq: Denominator}) can be reduced to a cubic polynomial of $y^2$, we can find the seven roots of $D_{x'}$, which we present in Appendix~\ref{App: Regularization} to avoid overly long expressions here. One of these roots is $y = 0$. Notably, the system may still enter the central zone through the pseudo-stationary states $U_\pm$ at $x = \pm 1$ and $y = 0$.

This can be understood by noting that the dynamical equation for $x$ can be \textit{regularized}, where both the numerator and denominator of $x'$ vanish simultaneously. The numerator, $N_{x'}$, can be written as:
\begin{equation}
    N_{x'} = \sum_{i=0}^{3} f_i(c_1, c_2, y) x^i,\label{eq: Numerator}
\end{equation}
where $f_i(c_1, c_2, y)$ are polynomials in $y$, which we present in Appendix~\ref{App: Regularization}. Once we determine $y$ from the roots of $D_{x'}$, the expression for $N_{x'}$ becomes a third-degree polynomial in $x$, whose roots determine the location in phase space of singularities, i.e., points where the system is indeterminate. For example, when $y = 0$, we find $N_{x'} = 2(x^2 - 1)$, whose roots are $x = \pm 1$. The analysis of other roots is more complex, so we focus on an illustrative case with specific initial conditions and parameters.

In Fig.~5(b) of Ref.~\cite{Garnica:2021fuu}, it is demonstrated that a given trajectory in phase space is unable to enter the central zone 
%as it reaches a region where the variable $z^4$ becomes negative. As we have shown, problems in the dynamical description arises not for $z^4$ to be negative, but for its flip of sign during the evolution. 
{if it reaches a point where $D_{x^\prime}$ vanishes.}
In this case, what happens is that the system finds a regularization point, i.e., a singularity. In this analysis, we determine the precise point where this indeterminacy arises. Following the setup from Fig.~5(b) in Ref.~\cite{Garnica:2021fuu}, we choose the initial conditions:
\begin{equation}
    x_i = -4 \times 10^9, \quad y_i = 10^{10},
\end{equation}
with parameters:
\begin{equation}
    \alpha_1 = 1, \quad \alpha_3 = 1.1, \quad \chi_5 = 0,
\end{equation}
which correspond to $c_1 = 0.2$ and $c_2 = 2.2$. Using Eq.~\eqref{Eq: Expected N_inf}, the inflationary phase is expected to last approximately 560 e-folds.

Solving for the roots of $D_{x'}$ gives one trivial solution $y = 0$ and six non-trivial solutions. After evaluating them for $c_1$ and $c_2$, we find two complex roots and four real roots: $y = \pm 0.479592$ and $y = \pm 0.194975$. Singularities occur at $y = \pm 0.479592$, the largest values of $|y|$, which will be firstly met when the trajectory comes from the attractor line. Solving for the roots of $N_{x'}$ at these values gives $x = \pm 0.653576$, yielding two singularities at the points:
\begin{align}
    S_+ &= \{-0.653576, 0.479592\}, \\
    S_- &= \{0.653576, -0.479592\}.
\end{align}

These results are confirmed numerically in Fig.~\ref{Fig: No Central Zone}, which shows two trajectories approaching the singularity points $S_\pm$ from the attractor line $y = \beta_0 x$. After approximately 553.4 e-folds, the system diverges at these singularities, preventing access to the central zone spanned by $U_\pm$. Thus, the central zone can only be accessed if the initial conditions start within it.

In summary, further singularities may arise in the system due to the existence of regularization points, where the dynamical equation for $x$ becomes indeterminate, as both its denominator $D_{x'}$ and its numerator $N_{x'}$ vanish simultaneously. A natural question then arises: if $U_\pm$ and $S_\pm$ are singularities resulting from the regularization of the autonomous system, why can one case be numerically integrated whereas the other cannot? The answer lies in the numerical evaluation of $x'$ during the integration process. Numerical solutions are computed in discrete steps, and singularities can sometimes be ``skipped'' if a sufficiently small neighborhood around the pseudo-fixed point is well approximated. This enables the system to progress through regions near singularities, making numerical integration feasible, while preventing the trajectory from fully reaching the singular points.

In the left panel of Fig.~\ref{Fig: Numerical singularity}, the phase space trajectory (using the same initial conditions and parameters as in Fig.~\ref{Fig: Example Phase Space}) follows a well-behaved evolution. It enters the central zone through $U_-$ and then escapes. The colour bar shows that as the trajectory approaches $U_-$ from positive $y$, $|x'|$ grows to large values, which could potentially cause numerical indeterminacy, marked by the white region.\footnote{The large numerical values shown on the plot correspond to the white regions where $|x^\prime| > 100$, reaching maximum values of $|x^\prime| \sim 10^{60}$. These values are too large to be handled by our numerical integration method.} However, the width of this white region is smaller enough than the numerical integration step, allowing some trajectories to bypass the singularity.

Conversely, the right panel of Fig.~\ref{Fig: Numerical singularity} shows the phase space trajectory (using the same initial conditions and parameters as in Fig.~\ref{Fig: No Central Zone}). In this case, a wider white region around the regularization point $S_+$ prevents continuous integration of the system, leading to an indeterminacy.

%%%%%%%%%%%%%%%%%%%%%%%%%
\section{Conclusions} 
\label{Sec: Conclusions}
%%%%%%%%%%%%%%%%%%%%%%%%%

%The GSU2P theory has recently garnered attention for its potential to implement various cosmic phenomena, such as primordial inflation~\cite{Garnica:2021fuu}, late-time inflation~\cite{Rodriguez:2017wkg}, the existence of particle-like solutions~\cite{Martinez:2022wsy}, as well as black hole~\cite{Gomez:2023wei} and neutron star solutions~\cite{Martinez:2024gsj}. However, the full cosmic history predicted by this model has not been thoroughly explored. Motivated by this gap, our work undertook a detailed investigation of the GSU2P theory embedded in a flat FLRW spacetime, using a cosmic triad configuration, aiming to identify the conditions under which it could consistently drive cosmic acceleration in line with observations. Our analysis, however, uncovered several instabilities that fundamentally challenge the GSU2P theory's viability in explaining cosmic acceleration. In the following, we summarize our key findings.

The GSU2P has been carefully studied through a dynamical system approach. We have demonstrated that the fixed points of the system fail to yield viable cosmological scenarios.  Specifically, the Hubble parameter is either infinite or indeterminate at these points or during the system's evolution, as observed in the case of the points $A_\pm$ on the right-hand side of Fig.~\ref{Fig: Central Zone}. This behaviour prevents the existence of stable accelerated attractors within the model. However, the chosen dynamical variables form a non-compact phase space, and additional stationary states could theoretically exist. We have identified two pseudo-stationary states. The first corresponds to three straight lines that dominate the system’s behaviour in the large $x$ and $y$ regime, where $x$ represents the velocity of the field and $y$ its magnitude. The second involves two points, $U_\pm$, which are solutions to the dynamical equation for $x$ when $y$ approaches zero.

In the large $x$ and $y$ regime, we found that one of the attractor lines aligns with a de-Sitter-like expansion consistent with constant-roll dynamics. In contrast, the pseudo-stationary points $U_\pm$ act as saddles, giving rise to what we have termed the ``central zone''. When trajectories enter this zone, oscillations between the pseudo-stationary states $U_\pm$ occur, mimicking a radiation-like behaviour of the field. However, these oscillations lead to instability, as the Hubble parameter becomes indeterminate at each crossing. After several oscillations, the system escapes the central zone, experiencing uncontrolled growth in the field magnitude, ultimately leading to a divergence. 
%This instability renders the GSU2P model unsuitable for both inflation and late-time cosmic acceleration, as we further explain below.

Regarding inflation, our findings indicate that the instability prevents a smooth transition from an inflationary phase to a radiation-dominated epoch, a process that would otherwise occur along the attractor line $y = \beta_0 x$. The instability arises from the indeterminacy of the Hubble parameter when the system crosses the points $U_\pm$, obstructing the graceful exit from inflation. Furthermore, the absence of limit cycles within the central zone exacerbates this instability, as trajectories rapidly escape the central zone, disrupting the radiation-dominated era. 

For late-time cosmic acceleration, the instability similarly undermines the model’s potential. The saddle-like behaviour of the points $U_\pm$ prevents the system from sustaining periodic oscillations, forcing the trajectories out of the central zone. Once outside, the lack of stable attractor points leads the trajectories to diverge, causing the vector field to quickly dominate the cosmic budget. This precludes the model from producing a viable cosmological history where a matter-dominated epoch is followed by a phase of accelerated expansion driven by the vector field.

In addition to these instabilities, we have also identified potential singularities in the phase space arising from the regularization of the dynamical equation for $x$. These singularities occur when both the numerator and denominator of the equation vanish simultaneously. Under certain conditions, these singularities prevent the system from entering the central zone. Our numerical simulations showed that trajectories following the attractor line can reach these singularities after a finite time, at which point the system becomes non-integrable, halting its evolution. These findings suggest that incomplete nullcline curves in the phase space introduce further instabilities into the system’s dynamics.

In conclusion, the GSU2P theory faces significant obstacles in providing a cosmologically viable explanation for both primordial inflation and late-time cosmic acceleration. Although the theory can reproduce a constant-roll phase, the numerous numerical instabilities—the absence of limit cycles, and the occurrence of singularities—render the system unable to maintain a viable cosmological evolution across cosmic timescales. 

Are these results sufficient to conclusively rule out the theory? Although not definitive, we can confidently state that reducing the dimension of the parameter space so that the theory behaves perturbatively (to second order in the tensor sector of the action) like GR and gives way to a non-anomalous gravitational wave speed, which represents the simplest realization, is cosmologically unviable. Other trivial parameter choices, though largely unexplored due to their mathematical complexity, could either mitigate or exacerbate the instabilities already observed, highlighting the need for further investigation. These results point to the need for further refinement of the theory to reconcile it with the known expansion history of the universe.

%%%%%%%%%%%%%%%%%%%%%%%%%%%%%
\section*{Acknowledgements}
%%%%%%%%%%%%%%%%%%%%%%%%%%%%%

This research work has been funded by  Universidad Antonio Nariño under Grant No. VCTI 2024211, by Universidad Industrial de Santander under the grant VIE 3921 (Universidad del Valle Grant No. 71388) and by Vicerrector\'ia de Investigaciones - Universidad del Valle Grants No. 71383 and 71373.

%%%%%%%%%%%
\appendix
%%%%%%%%%%%

%%%%%%%%%%%%%%%%%%%%%%%%%%%%%%%%%%%%%%%%%%%%%%%%%%%%%%%%%%%%%%%%%%%%%%%%%%%%%
\section{Dynamical Analysis of the Pseudo-Stationary Straight Lines.}
\label{App: Straight Lines}
%%%%%%%%%%%%%%%%%%%%%%%%%%%%%%%%%%%%%%%%%%%%%%%%%%%%%%%%%%%%%%%%%%%%%%%%%%%%%

In Section \ref{SEC: TSSL}, we have introduced the existence of additional straight lines with slopes \( \beta_\pm \) [Eq.~\eqref{Eq: beta+-}] that influence cosmological dynamics. Unlike the line with slope \( \beta_0 \), which exists independently of \( c_1 \) and \( c_2 \) values, the lines with slopes \( \beta_\pm \) only arise if \( (c_1 - c_2)(c_1 - 7c_2) \geq 0 \).

Each line can dominate the dynamics when it has attractor stability. Reference~\cite{Garnica:2021fuu} shows that small perturbations around any point $x_s$ on these lines maintain the attractor condition if
\[
    A_{\beta} = \frac{x_s}{x^{\prime}(x_s)} \frac{\partial x^{\prime}}{\partial x} \Big|_{x_s} > 1,
\]
where the attractor conditions for each line are found as
\begin{align}
    A_{\beta_0} &= -6 + \frac{7 c_2}{c_1},\\
    A_{\beta_{\pm}} &= 2 + 2 \frac{c_1}{c_2} \pm 2 \frac{ \sqrt{(c_1 - 7 c_2)(c_1 - c_2)}}{c_2}.
\end{align}

Thus, the line with slope \( \beta_0 \) is an attractor when \( |c_1| < |c_2| \). Similarly, with \( (c_1 - c_2)(c_1 - 7c_2) \geq 0 \), the line with slope \( \beta_{-} \) becomes an attractor if
\[
    c_2 < 0 \lor \left\{c_2 > 0 \land \left(\frac{3 c_2}{4} < c_1 \leq c_2 \lor c_1 \geq 7 c_2\right)\right\},
\]
whereas the line with slope \( \beta_+ \) is an attractor under
\[
    c_2 > 0 \lor \left\{c_2 < 0 \land \left(c_2 \leq c_1 < \frac{3 c_2}{4} \lor c_1 \leq 7 c_2\right)\right\}.
\]
For large values of \( x \) and \( y \), the system naturally evolves along these lines. However, the lines with slopes \( \beta_\pm \) do not inherently generate accelerated expansion; specific parameter adjustments are required to satisfy \( w_B < -1/3 \) [Eq.~\eqref{Eq: Equation of state beta}]. When all lines are present, stability analysis shows that each line can potentially act as an attractor depending on the values $c_1$ and $c_2$. Nonetheless, the dynamics in the central zone exhibit the same physical shortcomings as those previously identified for the line with slope \( \beta_0 \).

%%%%%%%%%%%%%%%%%%%%%%%%%%%%%%%%%%%%%%%%%%%%%%%%%%
\section{Long Expressions From Regularization}
\label{App: Regularization}
%%%%%%%%%%%%%%%%%%%%%%%%%%%%%%%%%%%%%%%%%%%%%%%%%%

Previously, in the Eq.~\eqref{Eq: Denominator}, we have described the denominator of $x^\prime$ as a seventh-degree polynomial in $y$. We present the roots of this polynomial below:
\begin{align}
    y_0&=0,\\
    y_1^2&=-\frac{\gamma}{3\delta}+\frac{\sqrt[3]{2} \Delta }{3 \delta  \sqrt[3]{\sqrt{\Xi ^2-4 \Delta ^3}-\Xi }}\\
    &+\frac{\sqrt[3]{\sqrt{\Xi ^2-4 \Delta ^3}-\Xi }}{3 \sqrt[3]{2} \delta },\notag\\
    y_2^2&=-\frac{\gamma }{3 \delta }-\frac{\left(1-i \sqrt{3}\right) \Delta }{3\ 2^{2/3} \delta  \sqrt[3]{\sqrt{\Xi ^2-4 \Delta ^3}-\Xi }}\\
    &-\frac{\left(1+i \sqrt{3}\right) \sqrt[3]{\sqrt{\Xi ^2-4 \Delta ^3}-\Xi }}{6 \sqrt[3]{2} \delta },\notag\\
    y_3^2&=-\frac{\gamma }{3 \delta }-\frac{\sqrt[3]{-2} \Delta }{3 \delta  \sqrt[3]{3 \sqrt{3} \sqrt{\zeta}-\Xi }}\\
    &-\frac{\left(1-i \sqrt{3}\right) \sqrt[3]{\sqrt{\Xi ^2-4 \Delta ^3}-\Xi }}{6 \sqrt[3]{2} \delta },\notag
\end{align}
where 
\begin{align}
    \Delta&\equiv 3 \alpha  \delta +\gamma ^2,\\
    \Xi&\equiv 9 \alpha  \gamma  \delta +2 \gamma ^3+27 \delta ^2,\\
    \zeta&\equiv\delta ^2 \Big[\left(2 \gamma -\alpha ^2\right) (\alpha  \delta +\Delta )+\alpha  \gamma  \delta +\Xi \Big].
\end{align}
Additionally, the numerator of $x^\prime$, given by Eq.~\eqref{eq: Numerator}, was described as a third degree polynomial on the $x$ variable where the functions $f_i\equiv f_i(c_1,c_2,  y)$ are given by:
\begin{align}
    f_0&=-2 \left(6 c_1 y^4+1\right) \left(12 c_2 y^4-y^2+1\right),\\
    f_1&=y \Bigg[3+12 y^2 \Big\{4 y^4 \left(2 c_1^2-4 c_1 c_2-7 c_2^2\right)\\
    &+y^2 (c_1+2 c_2)+2 c_1-5 c_2\Big\}\Bigg]\notag,\\
    f_2&=-2 \Bigg[8 y^6 \left(7 c_1^2-29 c_1 c_2+49 c_2^2\right)\\
    &+6 y^4 (c_1-4 c_2)+6 c_2 y^2-1\Bigg]\notag,\\
    f_3&=4 y^3 \Big[12 c_2 y^2 (4 c_1-7 c_2)-5 c_1+8 c_2\Big].
\end{align}

%%%%%%%%%%%%%%%%%%%%%%%%%%%%%%%%%%
\bibliographystyle{utcaps} 
\bibliography{Bibli.bib}
%%%%%%%%%%%%%%%%%%%%%%%%%%%%%%%%%%

\end{document}